\documentclass[twocolumn,letterpaper]{IEEEAerospaceCLS}  % only supports two-column, letterpaper format

\usepackage{rotating}
\usepackage[]{graphicx}    % We use this package in this document
\newcommand{\ignore}[1]{}  % {} empty inside = %% comment

\usepackage{amssymb}
\usepackage{url}

\begin{document}

\title{NOIRE Study Report: Towards a Low Frequency Radio Interferometer in Space}
\author{%
Baptiste Cecconi, Moustapha Dekkali, Carine Briand, Boris Segret\\
LESIA, Observatoire de Paris, PSL, CNRS, Sorbonne Universit\'es, UPMC, Univ.\ Paris Diderot, Sorbonne Paris Cit\'e\\
Meudon, France\\
+33 1 45 07 77 59 --- 
baptiste.cecconi@obspm.fr
\and
Julien N. G. Girard\\
CEA/AIM, Univ.\ Paris Diderot, Sorbonne Paris Cit\'e\\
Saclay, France\\
julien.girard@cea.fr
\and
Andr\'e Laurens, Alain Lamy, David Valat, Michel Delpech, \\
Mickael Bruno, Patrick G\'elard\\
CNES, Toulouse, France\\
+33 5 61 27 46 42 --- 
andre.laurens@cnes.fr
\and
Martin Bucher\\
APC, Univ.\ Paris Diderot, Sorbonne Paris Cit\'e\\
Paris, France
\and
Quentin Nenon\\
ONERA\\
Toulouse, France
\and
Jean-Mathias Grie{\ss}meier\\
LPC2E, Universit\'e d'Orl\'eans, CNRS\\
Station de Radioastronomie de Nan\c cay, Observatoire de Paris, PSL, CNRS, Univ. Orl\'{e}ans, OSUC\\
Orl\'eans, France
\and
Albert-Jan Boonstra\\
ASTRON\\
Dwingeloo, the Netherlands
\and 
Mark Bentum\\
Eindhoven Technical University,\\
Eindhoven, the Netherlands\\
\thanks{\footnotesize 978-1-5386-2014-4/18/$\$31.00$ \copyright2018 IEEE}              % This creates the copyright info that is the correct 2018 data.
}

\maketitle
\thispagestyle{plain}
\pagestyle{plain}

\begin{abstract}
Ground based low frequency radio interferometers have been developed in the last decade and are providing the scientific community with high quality observations. Conversely, current radioastronomy instruments in space have a poor angular resolution with single point observation systems. Improving the observation capabilities of the low frequency range (a few kHz to 100 MHz) requires to go to space and to set up a space based network of antenna that can be used as an interferometer. 
   
We present the outcome of the NOIRE (Nanosatellites pour un Observatoire Interf\'erom\'etrique Radio dans l'Espace / Nanosatellites for a Radio Interferometer Observatory in Space) study which assessed, with help of CNES' PASO\footnote{PASO - Architecture Platform for Orbital Systems - is CNES' cross-disciplinary team in charge of early mission and concept studies}, the feasibility of a swarm of nanosatellites dedicated to a low frequency radio observatory. With such a platform, space system engineering and instrument development must be studied as a whole: each node is a sensor and all sensors must be used together to obtain a measurement. The study was conducted on the following topics: system principle and concept (swarm, node homogeneity); Space and time management (ranging, clock synchronization); Orbitography (Moon orbit, Lagrange point options); Telecommunication (between nodes and with ground) and networking; Measurements and processing; Propulsion; Power; Electromagnetic compatibility. 

No strong show-stopper was identified during the preliminary study, although the concept is not yet ready. Several further studies and milestones are identified. The NOIRE team will collaborate with international teams to try and build this next generation of space systems.  
\end{abstract}

\section{Introduction}
Radioastronomy is a young astrophysical science. The first detection of extraterrestrial radio emission occurred in 1932 \cite{1933PA.....41..548J}. Despite such short history, radio observatories are now coming to a golden age, which is enabling new types of observations down to very low frequencies. Table \ref{tab:radio-ground-space} presents a selection of a few milestone and shows the evolution of the complexity and type of radio telescope: ground based telescopes have been gaining angular resolution by using array of antenna (either phased arrays, like UTR-2 or NDA, or interferometers, like LOFAR), whereas space observatories are still in the single antenna era. Space systems are required for observations below the ionospheric cut-off at $\sim$ 10 MHz. However, building instruments with angular resolution at those frequencies has long been impossible to implement, due to the required aperture size (the angular resolution of a telescope is $\sim\lambda/D$ with $\lambda$ the observed wavelength and $D$ the aperture size). Other techniques have been used to recover angular resolution with simple antennas, such as goniopolarimetric techniques \cite{lecacheux_AA_78,cecconi_RS_05}. With strong assumptions on the incoming wave (single point source at infinity), direction-finding and polarization with accuracy up to $\sim1^\circ$ and $10\%$ are reachable respectively with intense signals. The Cassini and STEREO missions are using these techniques and the research teams have published many results including radio source characterization (such as a posteriori radio source mapping). However, real imaging capabilities is still out of reach of current space radio instruments. 

During the last decade, several concept studies and projects (see Table \ref{tab:space-array-projects}) have been drafted with the aim of building a radio interferometer in space, and thus allows sky imaging down to a few kHz. The NOIRE study presented here is part is this international effort to increase our knowledge, to identify and improve the technologies that will lead us to a future low frequency radio observatory in space. The NOIRE study was initiated after regular discussions and contacts with the Dutch team at the origin of several studies and projects: OLFAR (Orbiting Low Frequency Array), DARIS (Distributed Aperture Array for Radio Astronomy in Space), DEx (Dark Ages Explorer), SURO (Space based Ultra long wavelength Radio Observatory), DSL (Discovering the Sky at the Longest Wavelengths) \cite{Saks:2010uf,Bentum:2011vh,Boonstra:2016tc}.

Once out of the ionosphere, observations of the sky is possible down to the local heliosphere radio wave cut-off frequency (about 10 kHz). However, the human activity is producing a radio noise (Radio Frequency Interferences, RFI), that pollutes the spectral range down to a few hundred kHz. Observations by RAE-B (Radio Astronomy Explorer-B) \cite{alexander_AA_75} or more recently with the Kaguya/LRS (Lunar Radar Sounder) \cite{Ono:2010ei} shows that the closest place near Earth free of RFI is the far side of the Moon. 

\begin{table*}
\centering\begin{tabular}{llccccc}
&&&\multicolumn{2}{c}{Observatory}&\multicolumn{2}{c}{Antenna}\\
Date&Milestone&Frequency(MHz)&Ground&Space&Single&Array\\\hline
1932&First observation of Galactic radio emission \cite{1933PA.....41..548J}&20.5 &$\times$&&$\times$&\\
1946&First interferometric radio astronomy measurement \cite{1946Natur.158..633P,1947RSPSA.190..357M}&200 &$\times$&&$\times$&\\
1955&Detection of Jovian radio emissions \cite{burke_JGR_55}&22.2 &$\times$&&$\times$&\\
1972&First light of UTR-2 (Kharkov, Ukraine) \cite{Konovalenko:2016gv}&$8 - 40$&$\times$&&&$\times$\\
1973&First map of Galactic background emission (RAE-B) \cite{novaco_ApJ_78}&$<13.1$&&$\times$&$\times$&\\
1977&First light of NDA (Nan\c cay Decameter Array, France) \cite{Boischot:1980vj}&$8 - 80$&$\times$&&&$\times$\\
1997&Launch of the Cassini mission (NASA) \cite{gurnett_SSR_04}&$<16$&&$\times$&$\times$&\\
2007&Launch of the STEREO mission (NASA) \cite{bougeret_SSR_08}&$<14$&&$\times$&$\times$&\\
2012&LOFAR radio telescope in Europe \cite{lofar}&$10-240$&$\times$&&&$\times$\\
2013&Long Wavelength Array 1 (LWA) \cite{lwa}&10 -- 88&$\times$&&&$\times$\\
\end{tabular}
\caption{Ground and space low frequency radio astronomy milestone}\label{tab:radio-ground-space}
\end{table*}

\begin{table*}
\centering\begin{tabular}{llllll}
Name&Frequency &Baseline&Nb of S/C&Location&Team / Country\\
&(MHz)&(km)&&&\\
\hline
SIRA&0.03 -- 15&$>$10&12 -- 16&Sun-Earth L1 halo&NASA/GSFC (2004)\\
SOLARA/SARA&0.1 -- 10&$<$10000&20&Earth-Moon L1&NASA/JPL - MIT (2012)\\
OLFAR&0.03 -- 30&$\sim$100&50&Lunar orbit&ASTRON/Delft, NL (2009)\\
     &          &            &  & or Sun-Earth L4-L5&\\ 
DARIS&1 -- 10&$<$100&9&Dynamic Solar Orbit&ASTRON/Nijmegen, NL\\
DEX&0.1 -- 80&$\sim$1&$10^5$&Sun-Earth L2&ESA-L2/L3 call\\
SURO&0.1 -- 30&$\sim$30&8&Sun-Earth L2&ESA M3 call\\
SULFRO&1 -- 100&$<$30&12&Sun-Earth L2&NL-FR-China (2012)\\
DSL&0.1 -- 50&$<$100 &8&Lunar Orbit (linear array)&ESA-S2 (2015)\\
DEX2&0.1 -- 80&100&10 -- 100&Lunar Array&ESA-M5 (2016)\\
SunRISE&0.1 -- 25&12&6&GEO graveyard&NASA concept study (2017)
\end{tabular}
\caption{Recent history of space-based radio observatory studies}
\label{tab:space-array-projects}
\end{table*}

The goal of the NOIRE study is provide an initial assessment of feasibility and reachable performances for a space-based radio observatory implemented on an homogeneous low-control swarm of spacecraft. Starting from the state of the art of already published studies and concepts, the NOIRE team has identified a selection of key science objectives (corresponding to the topical expertise of the science team) and conducted a feasibility study with the CNES/PASO team. The NOIRE team is presented in Appendix \ref{app:team}.

\section{Science Objectives}

In the low frequency range (namely below 100 MHz), the sky brightness temperature can be as high as $10^7$ K at about 1 MHz. Figure \ref{fig2} is showing the main radio sources and components observable in space near Earth. As discussed in previous studies \cite{Bentum:2011vh,Burns:2012gs,Zarka:2012ck,KleinWolt:2013wk,Boonstra:2016tc}, many science objectives are concerned by such instrumentation. We present here a few of them, focussing on the science expertise of the team:
\begin{itemize}
\item\textsl{Cosmology} --- Pathfinder measurements of the red-shifted HI line that originates from before the formation of the first stars (dark ages, recombination) \cite{Burns:2012gs}; 
\item\textsl{Astrophysics} --- Low frequency sky mapping and monitoring: radio galaxies, large scale structures (clusters with radio halos, cosmological filaments...), including polarization, down to a few 100 kHz \cite{novaco_ApJ_78}; 
\item\textsl{Pulsars} --- Radio observations at the lowest frequencies will allow to test the temporal broadening, which is usually assumed to vary with $f^{-4}$. In this context, it will be possible to determine whether there is a low-frequency cut-off for this broadening relation, which could be indicative of the largest turbulence scales in the interstellar medium. This will also allow to put constraints on the observation of radio transients a low frequencies;
\item\textsl{Inner heliosphere} --- Low-frequency radio bursts from the Sun, from 1.5 $R_S$ (Solar Radius) to $\sim$1 AU (Astronomical Unit) \cite{Krupar:2014ke,Morosan:2014ff}.
\item\textsl{Space Weather} --- Scintillation and Faraday rotation through the heliospheric plasma \cite{Genova:1981tv}. 
\item\textsl{Planetary Magnetospheres} --- Auroral emissions from the Earth and from the giant planets' magnetospheres in our solar system: rotation periods, modulations by satellites and Solar Wind, magnetospheric dynamics, seasonal effects... First opportunity in decades to study Uranus and Neptune \cite{zarka_ASR_92} Earth and Jupiter radiation belts \cite{Girard:2016gy}. 
\item\textsl{Planetary Atmospheres} --- Lightning induced radio pulses at Earth, Saturn, Uranus and Neptune \cite{zarka_PRE1_85,zarka_Nature_86,Konovalenko:2013ga}.
\item\textsl{Exoplanets} --- Exoplanetary auroral radio emissions \cite{Zarka:2012ck}
\end{itemize}

\subsection{Detailed science objectives}
Two science objectives are discussed here: the Dark Ages and Planetary radio emissions.

\subsubsection{The dark ages}
\begin{figure*}
\centering\includegraphics[width=0.9\linewidth]{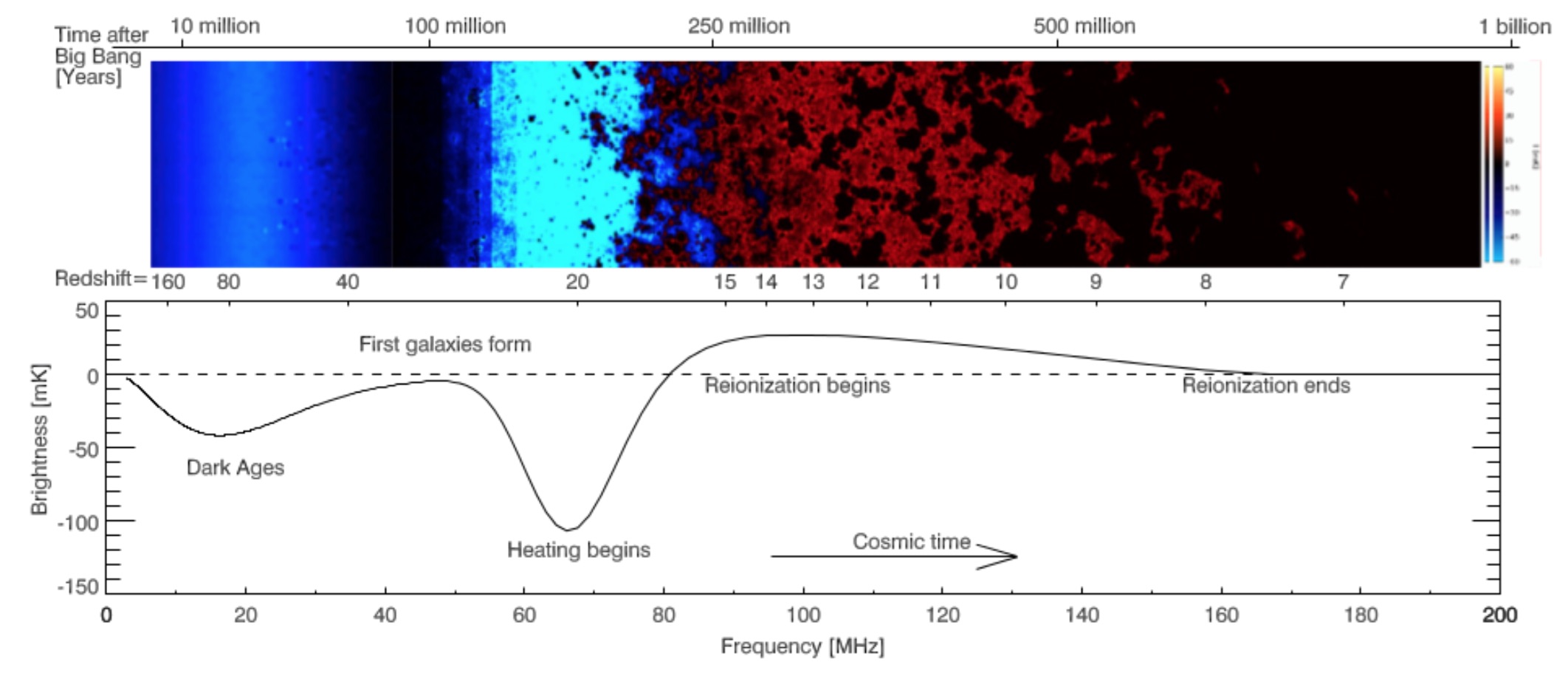}
\caption{Simulation of the Dark Ages signals (see \protect\cite{Pritchard:2012ja} for more details)}\label{fig:dark-ages}
\end{figure*}

The so-called Dark Ages signals are fossil signatures of the very early universe, from the reoinization to the forming of the first stars. This phenomenon is predicted to appear as spectral fluctuations between 5 and 80 MHz. Figure \ref{fig:dark-ages} is showing a simulation of this signature. Its amplitude is a few 10 mK as compared to the sky brightness temperature that can be as high as $10^7$ K at very low frequencies.

The observation of the Dark Ages is thus particularly difficult, as it is a relative measurement, compared to a basically unknown sky. The only knowledge of the sky is partial at low frequencies: some maps exist but are partial, not accurate, and difficult to reuse \cite{novaco_ApJ_78}. The decreasing sky brightness temperature helps the measurements at higher frequencies. Furthermore the anisotropy of the sky brightness temperature, is known but not characterized \cite{manning_AA_01}. 

If we assume a 30 mK signal at 30 MHz, with a 5 mK accuracy requirement for a 5$\sigma$ detection. At this frequency, the sky brightness temperature is about 1000 K. We need a measurement with $\Delta T/T \approx 5\times10^{-6}$. Assuming a 10 MHz spectral integration bandwidth, this would require
an integration time of 4000 seconds. This integration time shall be accumulated with data samples without any variable foreground source. Furthermore, this assumes that the instrument is perfectly calibrated and that this calibration is constant with time with a noise below the 5 mK accuracy requirement. 

\subsubsection{Planetary radio emissions}
\begin{figure*}
\centering\includegraphics[width=\textwidth,clip]{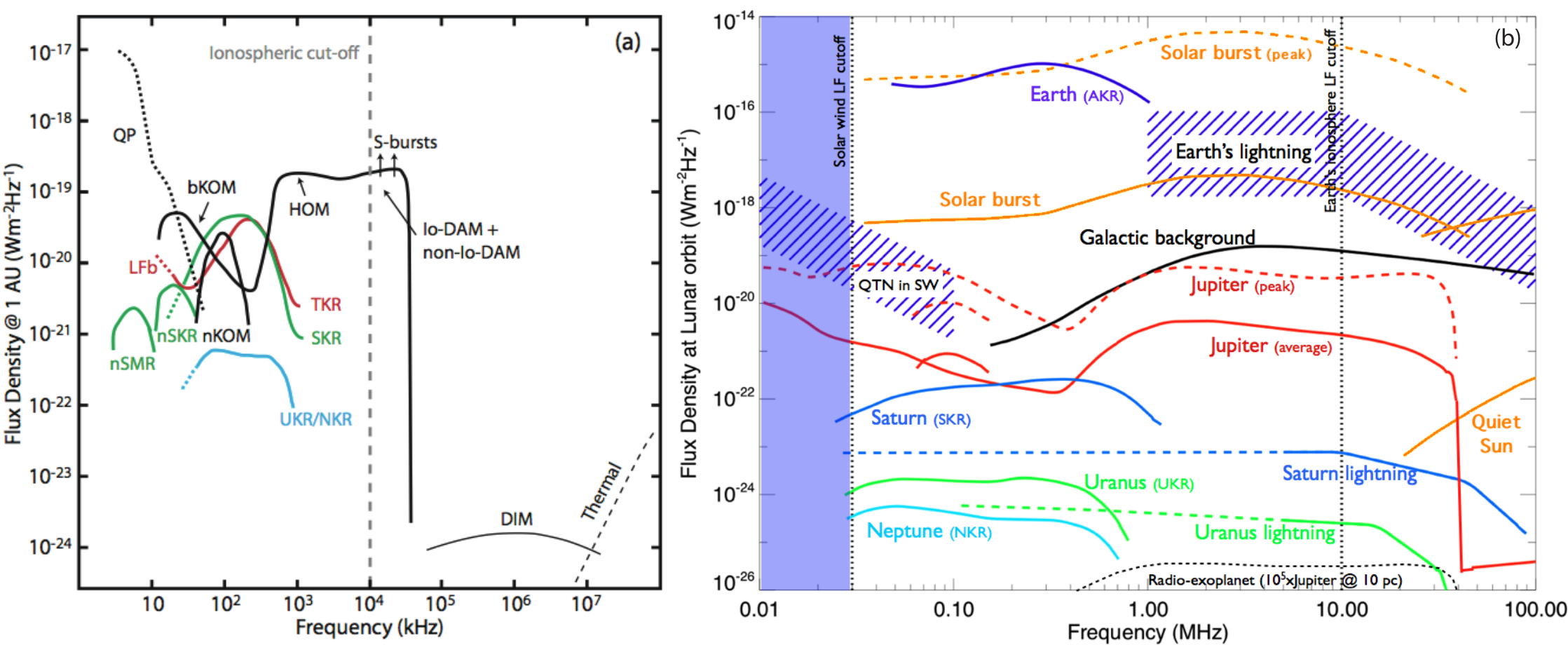}
\caption{(a) Solar system radio source normalized spectra, as observed from a distance of 1 AU (Astronomical Unit). Radio emission spectra from Jupiter, Saturn, Earth, Uranus/Neptune are traced in black, green, red and blue. (b) The radio component spectra observed from Earth, including solar radio emissions, earth and planetary auroral and atmospheric radio emissions, galactic emission and local plasma noise on the antenna. Figure adapted from \protect\cite{Zarka:2012ck} and \protect\cite{Cecconi:2014ih}.}
\label{fig2}
\end{figure*}

The magnetized planets, which are the Earth, Jupiter, Saturn, Uranus and Neptune in our solar system, are emitting intense radio signals resulting from particle accelerations in various places of the planetary environment. The auroral radio emissions are emitted above the magnetic poles of the planet through a cyclotron instability. The radiation belts (in the case of the Earth and Jupiter) are emitting synchrotron emission. The planetary atmospheric discharges can also be the source of radio pulses associated to lightnings. Figure \ref{fig2} shows these various components of the planetary electromagnetic spectra. Radio emissions can be used to remotely characterize accelerated particles on magnetized plasma, as well as study the dynamics of the planetary magnetic environment, or discharges in atmospheres. 

Except for Jovian radio emissions which are visible from ground observatories, most other components occur at too low frequency to be observed from ground. The current radio exploration is done with single point radio observatories (placed on planetary exploration space missions, such as Cassini, JUNO, JUICE...). Thanks to goniopolarimetric techniques, it is possible to reconstructed ``images'' after accumulating many single observations. However, this process is very limited and always assumes a single radio source per observation. The objective is to go from this ``direction of arrival'' era to direct imaging, as LOFAR allowed to do for the low frequency part of the Jovian radiation belts \cite{Girard:2016gy}. With an observatory located away from Earth, we could build images of the Earth radiation belts. 

\subsection{Science requirement analysis}
Table \ref{tab:sci-obj} lists the high-level science objectives identified by the NOIRE team and evaluated in this work. On that table, each high-level science objective is identified by its own S-ID. Each high-level objective is translated into ``Science Performance Requirement'' parameters. For some high-level science objectives, the team was not able yet to set the science performance requirements. The result of this preliminary analysis is presented in Table \ref{tab:sci-perf}. From this assessment, it is possible to classify the high-level science objective into ``Measurement Performance Requirements'', as presented in Table \ref{tab:meas-req}. Spectral (M-Fxx), temporal (M-Txx) and polarization (M-Pxx) measurement requirements are not binding for NOIRE, as they can be fulfilled with single spacecraft observatory. Signal (M-Sxx) and Spatial (M-Lxx) measurement requirements are used to define the NOIRE platform characteristics. The sensitivity is related to the number of nodes (the effective area of each antenna adds up). The spatial resolution is directly related to the length of the baselines (the lengths of the lines joining each pair of nodes). The requirements on the platform characteristics will be described in the next sections. 

%\begin{figure*}
%\centering\includegraphics[width=\textwidth,clip]{cecconi1_fig1.png}
%\caption{Cosmological science objectives for low frequency observations, adapted from \protect\cite{KleinWolt:2013wk}}
%\label{fig1}
%\end{figure*}

\begin{table*}
\centering\begin{tabular}{lllll}
S-ID&Science Topic&Observed Phenomena\\
\hline

\multicolumn{3}{l}{\textsl{Cosmology and Astrophysics}}\\
S-CA1&Low frequency anisotropy of CMB&21 cm Line redshifted to 5-30 MHz range\\
S-CA2&Foreground sources&Extragalactic sources\\
S-CA3&Foreground sources&Low frequency sky mapping\\
S-CA4&Pulsars&Low frequency dispersion of pulsars\\
%S-CA5&Transient events&Fast Radio Bursts\\ 
%S-CA6&Transient events&Radio Counter part of gravitational waves\\ 
\hline

\multicolumn{3}{l}{\textsl{\textsl{Solar and Stellar Physics}}}\\
S-SO1&Solar physics and Space Weather&Radio bursts associated with solar flares\\
S-SO2&Solar physics and Space Weather&Radio bursts associated with CME and interplanetary shocks\\
S-SO3&Solar physics and Space Weather&In-situ electrostatic waves\\	
S-SO4&Solar physics and Space Weather&Quasi thermal noise spectroscopy\\
S-ST1&Stellar physics&Stellar radio bursts\\
\hline

\multicolumn{3}{l}{\textsl{\textsl{Planetary and Magnetospheric Sciences}}}\\
S-PM1&Magnetospheric radio emissions&Terrestrial magnetospheric radio emissions\\
S-PM2&Magnetospheric radio emissions&Jovian magnetospheric radio emissions\\
S-PM3&Magnetospheric radio emissions&Kronian magnetospheric radio emissions\\
S-PM4&Magnetospheric radio emissions&Uranus and Neptune auroral radio emissions\\
S-PM5&Magnetospheric radio emissions&Exoplanetary auroral radio emissions\\
S-PA1&Planetary atmospheric electricity&Terrestrial lightnings\\
S-PA2&Planetary atmospheric electricity&Kronian lightnings\\
S-PA3&Planetary atmospheric electricity&Uranus lightnings\\
S-PB1&Planetary Radiation Belts&Earth radiation belts\\
S-PB2&Planetary Radiation Belts&Jupiter radiation belts
\end{tabular}
\caption{Science Objectives}
\label{tab:sci-obj}
\end{table*}

\begin{sidewaystable}
\centering\begin{tabular}{c|lll|lll|ll|lll|ll}
&\multicolumn{3}{c|}{Spectral Scale}&\multicolumn{3}{c|}{Signal Scale}&\multicolumn{2}{c|}{Spatial Scale}&\multicolumn{3}{c|}{Temporal Scale}&\multicolumn{2}{c}{Polarization}\\
S-ID&Min.&Max.&Resol.&Fluct.&Max.&Dyn.&Min&Max&Dur.&Repet.&Resol.&Circ.&Lin.\\
\hline
S-CA1&5 MHz&50 MHz&1 MHz&noise&1 Jy&70 dB&$4\pi$ sr&$4\pi$ sr&$\infty$&--&--&?&?\\
S-CA2&300 kHz&100 MHz&?&?&?&?&?&?&?&?&?&?&?\\
S-CA3&300 kHz&100 MHz&100 kHz&$10^{-3}$ Jy&$10^6$ Jy&90 dB&1'&$4\pi$ sr&?&?&?&?&?\\
S-CA4&1 MHz&100 MHz&100 kHz&noise&1 Jy&120 dB&\multicolumn{2}{c|}{unresolved}&100 ms&1 s&1 ms&yes&yes\\
S-ST1&300 kHz&100 MHz&?&?&?&?&?&?&?&?&?&?&?\\
S-SO1&10 kHz&100 MHz&10 kHz&noise&$10^{12}$ Jy&80 dB&1'&$90^\circ$&2 h&random&10 ms&\multicolumn{2}{c}{no below 20 MHz}\\
S-SO2&100 Hz&100 kHz&--&1 mV/m&200 mV/m&30 dB&--&--&10 ms&&10 $\mu$s&\multicolumn{2}{c}{electrostatic}\\
S-SO3&1 kHz&100 kHz&1\%&$10^6$ Jy&$10^8$ Jy&80 dB&--&$\infty$& &1 s& &0\%&100\%\\
S-PM1&10 kHz&1 MHz&1 kHz&noise&$10^{11}$ Jy&120 dB& & &10 h&24 h&1 ms&100 \%&0\%\\
S-PM2&10 kHz&45 MHz&1-1000 kHz&noise&$10^8$ Jy&120 dB& & &9.5 h&9.5 h&1 ms&100-70 \%&0-50\%\\
S-PM3&10 kHz&1 MHz&1 kHz&noise&$10^5$ Jy&120 dB& & &6 h&10 h&1 ms&100-70 \%&0-50\%\\
S-PM4&10 kHz&1 MHz&1 kHz&noise&100 Jy&120 dB& & &&&1 ms&&\\
S-PM5&300 kHz&100 MHz&10-1000 kHz&noise&1 Jy&120 dB&\multicolumn{2}{c|}{unresolved}&?&?&?&?&?\\
S-PA1&5 MHz&50 MHz&100 kHz&noise&$10^{10}$ Jy&100 dB& &1'&$<$1 ms&random&1 ms&&\\
S-PA2&500 kHz&50 MHz&100 kHz&noise&$10^3$ Jy&100 dB& &1'&$<$1 ms&random&1 ms&&\\
S-PA3&500 kHz&50 MHz&100 kHz&noise&100 Jy&100 dB& &1'&$<$1 ms&random&1 ms&&\\
S-PB1&100 kHz&1 MHz&10 kHz&$10^{-1}$ Jy&$10^{-3}$ Jy&60 dB& &10'&\multicolumn{2}{c}{continous}&15 min&&\\
S-PB2&10 MHz&100 MHz&1 MHz&noise&6 Jy&60 dB& &10''&\multicolumn{2}{c}{continous}&30 min&1\%&10\%
\end{tabular}
\caption{Science Performance Requirements}
\label{tab:sci-perf}
\end{sidewaystable}

\begin{table*}
\centering\begin{tabular}{cll}
M-ID&Description&S-IDs\\
\hline
     &\textsl{Perform a measurement of the radio spectrum:}&\\
M-F00&-- from 0.3 to 50 MHz with 10 kHz to 1 MHz resolution  &S-CA1, S-CA2, S-CA3, S-CA4, S-ST1, S-SO1,\\
     &                                                       &S-PM2, S-PM5, S-PA1, S-PA2, S-PA3, S-PB2\\
M-F01&-- from 50 to 100 MHz with $>$1 Mhz resolution         &S-CA2, S-CA3, S-CA4, S-ST1, S-SO1, S-PM5,\\
     &                                                       &S-PB2\\
M-F03&-- from 10 kHz to 1 MHz with $<$10 kHz resolution      &S-SO1, S-PM1, S-PM2, S-PM3, S-PM4, S-PB1\\
M-F04&-- from 100 Hz to 100 kHz with 1\% relative resolution &S-SO3\\
\hline
     &\textsl{Achieve a temporal sampling of the radio spectrum:}&\\ 
M-T01&-- with a temporal resolution $\approx$1 ms            &S-CA4, S-PM1, S-PM2, S-PM3, S-PM4, S-PA1,\\
     &                                                       &S-PA2, S-PA3\\
M-T02&-- with a temporal resolution $\gg$1 ms                &S-CA1, S-CA3, S-SO3, S-PB1, S-PB2\\
     &\textsl{Measure radio spectra over:}&\\ 
M-T10&-- the whole duration of the mission (persistent signal) &S-CA1\\ 
M-T11&-- periods of $\sim$1 h to 10 h 					     &S-SO1, S-PM1, S-PM2, S-PM3\\ 
M-T10&-- periods of $\sim$10 h to 30 h                       &S-PM1, S-PB1, S-PB2\\ 
M-T10&-- periods of $\sim$1 min to 1h                        &S-CA4, S-SO2, S-SO3, S-PM1, S-PM2, S-PA1,\\
     &                                                       &S-PA2, S-PA3\\ 
     &\textsl{Record a waveform of the electric field:}&\\ 
M-T20&-- with a sample every 10 $\mu$s for 10 ms             &S-SO2\\ 
\hline
     &\textsl{Measure radio flux density above galactic background:}&\\
M-S00&-- with signal level within $10^8$ to $10^{12}$ Jy     &S-SO1, S-SO3, S-PM1, S-PM2, S-PA1\\
M-S01&-- with signal level within $10^4$ to $10^8$ Jy        &S-CA1, S-CA3, S-PM3\\
M-S02&-- with signal level within $1$ to $10^4$ Jy           &S-CA4, S-PM4, S-PM5, S-PA2, S-PA3, S-PB2\\
M-S03&-- with signal level $<$1 Jy                           &S-CA1, S-CA4, S-PM5, S-PB1\\
     &\textsl{Measure fluctuating radio signals:}&\\
M-S10&-- with a dynamical Range $<$80 dB                     &S-CA1, S-SO1, S-SO2, S-SO3, S-PB1, S-PB2\\
M-S11&-- with a dynamical Range 100 dB                       &S-CA3, S-PA1, S-PA2, S-PA3\\
M-S12&-- with a dynamical Range 120 dB                       &S-CA4, S-PM1, S-PM2, S-PM3, S-PM4, S-PM5\\
     &\textsl{Measure electric field:}&\\
M-S20&-- with 1mV/m accuracy up to 200mV/m                   &S-SO2\\
\hline
     &\textsl{Spatially resolve the radio source:}&\\
M-L00&-- with a spatial resolution $>$1' on the full sky     &S-CA2, S-SO1\\
M-L01&-- with a spatial resolution $<$1' on 10' region       &S-PM1, S-PB1\\
M-L02&-- with a spatial resolution $<$1'' on 10'' region     &S-PM2, S-PB2\\
     &\textsl{Measure full sky signals:}&\\
M-L10&-- integrate on full sky (no spatial resolution)       &S-CA1, S-CA3, S-SO2, S-SO3\\
     &\textsl{Measure unresolved signals:}&\\
M-L20&-- at max resolution of instrument                     &S-CA2, S-CA4, S-ST1, S-PM3, S-PM4, S-PM5,\\
     &                                                       &S-PA1, S-PA2, S-PA3\\
\hline
     &\textsl{Measure signals without polarization:}&\\
M-P00&-- with no polarization                                &S-SO1 (below 20 MHz)\\
     &\textsl{Measure signals with polarization:}&\\
M-P10&-- with circular polarization with accuracy of 10\%    &S-CA4, S-PM1, S-PM2, S-PM3, S-PM4, S-PM5\\
M-P11&-- with circular polarization with accuracy of 1\%     &S-PB2\\
M-P12&-- with linear polarization with accuracy of 10\%      &S-CA4, S-SO3, S-PM2, S-PM3, S-PB2\\
\end{tabular}
\caption{Measurement requirements with corresponding science objectives.}
\label{tab:meas-req}
\end{table*}

%\begin{table*}
%\centering\begin{tabular}{clll}
%I-ID&Description&S-IDs&M-IDs\\
%%\hline
%%I-MO0&Waveform&S-SO2\\
%%I-MO1&Incoherent sum&S-CA1, S-SO3\\
%%I-MO2&Imaging&S-CA2, S-CA3, S-SO1, S-PM1, S-PM2, S-PM3, S-PM4, S-PB1, S-PB2\\
%%I-MO3&Beam-Forming&S-CA1, S-CA2, S-ST1, S-PM4, S-PM5, S-PA1, S-PA2, S-PA3\\
%\hline
%I-SF0&Spectral sampling step $<$10 kHz&&M-F03, M-F04\\
%I-SF1&Spectral sampling step $\approx$100 kHz&&M-F00\\
%I-SF2&Spectral sampling step $>$1 MHz&&M-F02\\
%I-ST0&Temporal sampling step $<$0.1 s&&\\
%I-ST1&Temporal sampling step $\approx$1 s&&\\
%I-ST2&Temporal sampling step $>$10 s&&\\
%I-SL0&Spatial sampling step $<$10' s&&\\
%I-SL1&Spatial sampling step $\approx 1^\circ$&&\\
%I-SL2&Spatial sampling step $>10^\circ$&&\\
%\hline
%I-IF0&Spectral integration step $<$10 kHz&&M-F22, M-F23, M-F23\\
%I-IF1&Spectral integration step $\approx$100 kHz&&M-F21\\
%I-IF2&Spectral integration step $>$1 MHz&&M-F20\\
%I-IT0&Temporal integration step $<$0.1 s&&\\
%I-IT1&Temporal integration step $\approx$1 s&&\\
%I-IT2&Temporal integration step $>$10 s&&\\
%I-IL0&Spatial integration step $<$10' s&&\\
%I-IL1&Spatial integration step $\approx 1^\circ$&&\\
%I-IL2&Spatial integration step $>10^\circ$&&\\
%\hline
%I-SS0&Sensitivity $<$1 mJy&&\\
%I-SS1&Sensitivity $<$0.1 Jy&&\\
%I-SS2&Sensitivity $<$1 Jy&&\\
%\end{tabular}
%\caption{Instrumental Capabilities vs. Measurement Performance Requirements}
%\label{tab:inst-perf}
%\end{table*}

\section{Principles and Concepts}

The aim of the NOIRE study is to assess on a system-wide point of view, the feasibility of a novel type of space instrumentation. The most innovative, disruptive, or even unconventional concepts have been preferred in order to face all major challenges and identify the potential hard spots. The study has been conducted in order to either adopt the proposed concepts (together with realistic implementations), or adjust or reject them if the implementation turned out to be unfeasible at this stage.

\subsection{Measurement Concepts}

Measurements performed with this new cutting-edge instrument take root in ground-based radio phased arrays and interferometers. Similar concepts therefore need to be adapted to the scope of current nanosatellites technology used in swarm. The added difficulty is to tackle the challenges raised by these 3D, potentially-deformable, array of sensors. As for other radio observatories, some measurements take place at the sensor level. These measurements are subsequently combined (in a distributed fashion) to form the array response, which is meant to bring improved angular resolution and sensitivity while dealing with individual sensors of manageable size and load.

Modern radio astronomy has entered a new golden age with the advent of continental-scale radio observatories such a LOFAR \cite{lofar} and SKA \cite{ska_2009}. Such observatories come with improved angular, spectral and temporal resolutions and huge sensitivity in a large field of view (FoV). Such improvements therefore come with a more complex instrumental ``calibrability''. Some critical assumptions, valid for ground based observatories, fall when going to space. For example, the lack of a (infinite) reflective ground plane (typically Earth ground) can introduce degeneracies in the calibration and imaging process.

Calibration of data taken on a large FoV demands the consideration of the so-called ``Direction-Dependent Effects'' (DDE) such as the strong variations of the antenna beam pattern with the direction. If not accounted for, the observed sky will strongly be biased by instrumental effect. On Earth, one can ignore this issue if the observed FoV is small enough to apply classical 2D small-field approximations. From space, such DDEs arise naturally by construction and the data processing should include such effects. Hopefully, the mathematical framework of the ``Measurement Equation'' (see \cite{Smirnov_2011a,Smirnov_2011b,Smirnov_2011c,Smirnov_2011d} and references therein) helps to model and implements methods to account for such effects.

Depending on the observing modes, the study requires to discuss the feasibility of using NOIRE for its various science cases.

\subsubsection{Individual Measurements}

Antenna Electro-magnetic (EM) simulations in vacuum can accurately represent the radiation properties of simple dipoles. However, attached to a nanosatellite body, the effective measured $\vec{E}$ field will corresponds to the projection of the incoming $\vec{E}$ field onto an ``effective'' antenna which direction depends on the coupling of the dipole with the satellite body. The resulting effective direction is hard to predict without refined EM simulations of the whole sensor. The precise knowledge of these effective directions are vital for the full 3D-polarimetric measurements, direction-finding and local measurements. In a swarm of monitored sensors that are all presenting various attitudes compared to absolute reference directions, we can transform all individual dipole orientation onto a single common reference frame gathering effective dipole directions, using linear transforms.  The precise localization and orientation of sensors with respect to this common reference frame requires an accurate bootstrapping on main absolute directions (generally provided by star trackers). Due to the symmetry of the sensors and that of the sky, there is not such a priori natural privileged direction.

Goniopolarimetric (GP) technique, or Direction-Finding, as it was used on Cassini until 2017, is based on deriving the direction of arrival (DoA) of a plane wave and its polarization, by using the correlation of signal amplitudes measured on the different available antenna polarizations. On this principle, GP is equivalent to having an ``angular'' spatial frequency sampling device, whereas a multiple antenna interferometer is a linear spatial frequency sampling device. Nevertheless, GP relies on strong assumptions on the regularity of the dipole beam pattern. In the short dipole regime ($l\ll\lambda/10$), the individual dipole beam pattern varies smoothly with direction following a $\sin^2$ function. When computing co-localized ``co-polar'' and ``cross-polar'' correlations on a single node between the $X$, $Y$ and $Z$ dipoles, one can derive analytical solutions to trace the orientation of an incoming wave vector $\vec{k}$. However, at higher frequencies, the beam pattern will present multiple nulls and many lobes that breaks the assumptions of GP \cite{cecconi_RS_05} and introduce as many degenerated directions that can fit the data. Hopefully, modeling antenna beam patterns in all direction as a DDE and using its spectral dependency, can help alleviate this issue.

In addition, improved knowledge of the antenna electric properties is critical for in-situ measurements of the surrounding plasma characteristics, such as thermal noise spectroscopy \cite{meyer_JGR_89}.

\subsubsection{Distributed Measurements}

The power of using a network of sensors resides in combining a lot of small manageable sensors to mimic the properties of a much larger instrument normally impossible to build nor launch but having improved angular resolution and sensitivity. This procedure is called ``aperture synthesis'' that is nowadays the default design approach to large scale instrument (e.g., LOFAR, MWA, SKA1-Low). The purpose of the study is to address its feasibility in space. In principle, there is no fundamental difference between building an array of radio antennas on the ground or in space. However, data rate constraints, online/offline data processing and orbitography, sensors relative and absolute positioning/attitude control are amongst the factors that will directly impede the theoretical performances of the array and the final scientific return of the measurements.

We need to distinguish between two main kinds of arrays: phased arrays and interferometers. While the two are proposing the reconstruction of an aperture in a equivalent manner (in the aperture plane or in the Fourier transform of the aperture plane), the computing loads and the format of the output data are drastically different. Beamforming is performed with a complex weigthed (amplitude and phase) sum that generate a single sensitivity beam pattern in a target direction. In our case, the signals measured along the main direction in the common reference frame of effective antennas.
An interferometer is a set of spatial frequency filters that samples the Fourier components of the sky (rather than the sky itself). In a complex-valued system, the sampled spatial frequencies depends on the array configuration. The measured samples are of the complex visibility function, and are called ``visibilities''. Practically, visibilities are obtained by computing the cross-correlation between two sensors signals.
 The computation load of the phased output is negligible compared to that of the correlator output. It can easily be affected by a misbehaving sensors that will corrupt the sum. Conversely, the computation of all correlations scales with $N^3$ per polarization where $N$ is the number of sensors. A misbehaving sensor can then be identified by its N-1 correlations (or baselines) and flag as bad values from the data.
 The two kind of arrays have similar pointing capabilities.
 
LOFAR is a typical working example of a fully digital radio telescope where the receiving parts are never moving. Beam forming, beam pointing are fully operated digitally. If we want to develop a N=50 network in space, we need to implement the intelligence that will take care of this operations, while the array is moving.

The digital pathway of the signal can be represented as flowing through a pipe of a certain radius. This radius depends on the maximum data transfer rate between the sensors and the processing unit as well as the processing capabilities of this unit. NOIRE will provide data distributed across frequency channels (see Fig. \ref{indivmeas}) and depending on the scientific case of observation. All frequencies can be allocated to a single particular direction of interest, or it can be subject of a trade-off between the number of desired directions and the bandwidth allocated per directions.

Regardless of the beamforming mode or interferometric mode, the default strategy will be to insert relative phase delay between the nodes to point the array in specific direction (i.e., the phase center). Coherency will be obtained in a certain angular region around the phase center. 
For the default targeted mode, the desired final imaging performance will have a strong influence on the design of the array. Working interferometers rely on the concept of coherency which guarantees that the sampled interferometric data is correctly capturing some information about the sky. This array will generate a lot of data, which can be reduced by spectral and temporal integration. However, it is known that such frequency or time averaging (in the correlator or during data post-processing) will affect the span of the field of view of a single pointing.
Indeed, by following classical radio textbooks (see \cite{condon_2016} Sect. 3.7.3), interferometry presents rule of thumbs that could help designing the instrument. In particular, for a single pointing at frequency $\nu_0$, the angular region $\Delta \theta$ for which we can expect to have sensible information at the resolution $\theta_r$ on the sky is linked to the frequency averaging windows $\Delta \nu$ thought the following inequality:
$\Delta \nu \Delta \theta \propto \theta_r \nu_0$
Likewise, time integration over a window of $\Delta t$ can affect the usable FoV through:
$\Delta t \Delta \theta \propto \theta_r \nu_0 P_{orb} / 2\pi$ where $P_{orb}$ is the effective rotation (or orbital) period of the array.

Depending on the choice and constraints put on the correlator, one the one hand, and the location and distribution of the array on the other hand, we can know what will be the angular size of each pointing. The total number of pointing to achieve a $4\pi$ will be $N_p=4\pi / \Omega_{p}$ where $\Omega_{p}$ is the solid angle subtended by the angular area of radius $\Delta \theta$.
The antenna, LNA and receiver systems will provide a measurement of the sky brightness with a certain variance. Combined together (in beamformed or interferometer mode), the measurements from the various sensors can be integrated to provide a certain SNR. For a fixed SNR, we can derive the survey speed (or the time to spend at each pointing) that is required if one wants to map the full sky at a certain depth. This figure of merit needs to be computed for the various possible array configurations, antennas, LNA but also for different time/frequency integration windows. The Full-sky map therefore need faceting imaging and mosaicking.

During the future design studies, we will push to propose a new original approach of imaging the sky by implementing an all-sky imaging mode. The feasibility of a cost-effective and efficient full-sky imaging mode, as well as the inversion methods and the required technologies to implement, are still under development. However, the interferometric problem needs to be recast in full 3D with a non-coplanar array \cite{cornwell_1992}. Nowadays computing power now allows for improved approach and calculation of a 3D Fourier transform. For the full-sky 3D approach on the sphere, one can still question the feasibility of computing and interpreting correlations on the whole sphere, and what information they contain about the sky.
The CHIMES experiment has demonstrated that with intensity mapping, it was possible to probe spherical harmonic modes of the sky by using specific slew observation strategies \cite{Liu_2016}. Sparse representation, in the scope of compressed sensing \cite{candes_2008} applied to radio interferometry (see for example \cite{garsden_2015,girard_2015}) can lead to robust representation of a signal on a sphere. Especially, spherical wavelets \cite{Lanusse_2012} can be a adapted dictionary to represent the sky, but still has to be linked to the measurement on full-sky.

The far field mapping of the sky is a primary goal of the instrument. Nonetheless, we can explore new reconstruction techniques that could allow tomographic studies of nearby sources such as Coronal Mass Ejection (CME) responsible for solar type III bursts seen in radio. Using a 3D interferometer, holography measurement become possible.

\subsubsection{Impact on System Specifications}

Array configuration and the number of sensors will depend on the tractability of computing $\sim N^2$ correlations but also the feasibility of controlling many sensors in the array.
For an RF array deployed on the Moon \cite{Jester_2009}, it is possible to derive the minimal number of sensors required to achieve a certain point source RMS. A typical number of 50 sensors will provide 1 Jy/beam, 100 mJy/beam, 10 mJy/beam point-source RMS in 24h respectively at frequencies of 3 -- 30 -- 300 MHz. The point-source RMS decreases with the number of sensors following a power law.
It is wrong to think that we will make a huge interferometer with million km baselines at low frequency n the frequency range of NOIRE ($f \leq 100$ MHz).
Being free from Earth ionospheric cut-off frequency (around 10 MHz), the array is still exposed to propagation effects in the interstellar medium (ISM) and interplanetary medium (IPM). their effect will be to restrict the maximal angular resolution power as a function of frequency. For example, the ISM and IPM angular broadening affect the size of any structures at 1 MHz to $\sim 1^\circ$ minimal angular size, which equates the maximum achievable angular resolution of an 50 km baseline at the same frequency. Temporal broadening will also distort the duration of transient radiosources. Detailed effects of the IPM, ISM, polarization decorrelation, absorption, angular/temporal dispersion and proximity of the sun and of the the galactic plane are developed in \cite{oberoi_2005}.

Integrating data helps improving the detection SNR but is limited to a bedrock defined as the thermal noise limit and the confusion limit. The confusion limit is the steady background radio noise associated with unresolved sources that create a structured background. Once this limit is achieved, there is no need to integrate further as the noise will stop decreasing. Depending on the maximum, this will limit the survey speed and the maximum achievable absolute sensitivity of the instrument. Confusion limit can be improved by increasing the maximum baseline and the number of sensors \cite{Jester_2009}, however, due to ISM/IPM scattering, at low frequencies, there is a fundamental limit to the detection level achievable by the instrument.

In order to be able to combine the raw signals into imaging products (see previous paragraphs), the platform must provide the following characteristics: 
\begin{itemize}
\item ranging accuracy on baselines: $\sim$ 1 cm (fraction of the shortest wavelength);
\item attitude of each node: better then 1$^\circ$ (accuracy of single node antenna calibration);
\item orientation of the baselines in an inertial astrophysical frame with an accuracy of $\sim$1'';
\item clock error or drift $<$5 ns (for 200 Msamples/s sampling rate) during the length of an individual snapshot;
\item the data must be sent to all computing nodes;
\item the orientation and length of all baselines at the time of the individual snapshots must be available on all computing nodes. 
\end{itemize} 
Linking the measurement performance requirement to the instrument and system performance requirement is not fully done yet because it requires further coupled modeling of the platform and measurement pipeline.

\subsection{System Architecture}

\subsubsection{Homogeneous Swarm}
All nodes of the swarm are identical, and convey the same hardware. The functions implemented are the same on each node, and may not however be activated at the same time for the same purpose. Each node can acquire signal and process it. All nodes can communicate with each other and with Earth. The interchangeability makes the swarm robust to failure. It also lowers the manufacturing costs. 

In case a fully homogeneous swarm appears to be not feasible, application-specific nodes would be envisaged for the such functions as space-time reference system, Earth communication, science processing unit.

\subsubsection{Networking}
The plan is to strictly mimic a ground-based network model with: physical connectivity with closest neighbors (not with all nodes); this connectivity doesn't need to be permanent but should be able to be activated at any time; logical connectivity between each nodes through Internet technologies (routing, protocols and services). 

Such a networking scheme allows a limited operating range for physical connectivity (and thus limited dedicated resources) and provides a full support to all communications needs: swarm measurement configuration (phasing of sensors); time reference propagation; shared computing; data distribution before and after the processing. 

\subsubsection{On-board Distributed Computing}

The usual strategy of transmitting a digitized waveform directly to Earth for processing, is probably not feasible within the low on-board resources of nanosatellites. Thus a good way to reduce significantly the data flow is to achieve the beamforming or interferometry processing on-board (even partially). For this purpose, it could be taken advantage of the total computing power available in the swarm to distribute this processing onto a networked computer architecture, inspired by its ground counterparts.

\subsubsection{Relative Navigation and Ranging -- Time Keeping and Propagation}

In order to achieve on-board interferometry processing, the individual time and position of each node must be known to an accuracy dictated by the observed phenomena, observation frequency, measurement requirements. Moreover, the individual measurements should be done within a given time slot for the correlation process to be acceptable.

For that purpose, an auto-location capability is necessary to estimate on-board the relative positions of the nodes, as well as the shape and orientation of the swarm. An on-board time reference is also required for the purpose of the interferometric processing (but also of the auto-location mechanism), to be maintained and distributed over the network. Such a time reference is not necessarily "absolute", i.e., locked on a terrestrial or astronomical time scale.

\subsubsection{Low Control Philosophy}

The swarm will fly with as little control as possible. The natural orbital evolution of the swarm is used instead of trying to strictly control and enforce its shape and the location of each of its nodes. 

Starting with the idea of a single step deployment (all nanosatellites dispensed from the same launcher), the initial state consists in having all nodes at the same place. Each node is ejected using standard spring dispensers, one at a time, possibly with attitude changes of the carrier. If we can achieve a swarm configuration suitable for measurements with solely natural evolutions (initial velocities, gravitational forces) from this initial conditions, there would be no need for dedicated post-dispersion operations. We can also imagine strategies such that measurements requiring short baseline are done early in the mission whereas longer baseline ones are done after the swarm has scattered. If a specific geometry is needed, then post-dispersion operations may be required. The long term evolution of the swarm geometry will have te be evaluate together with the science requirements. With the frame of such a low-control swarm, there may be a need to implement a herding dog algorithm \cite{hauert:2011}. The homogeneous swarm concept makes all nodes interchangeable. It is then possible to redistribute roles between central and remote nodes. This relaxes geometry constraints to statistical geometrical properties (contrarily to individual constraints, depending on node capabilities).

The measurement requirements don't impose any attitude control, but requires the knowledge of the attitude of each node, and a slow evolution of the node orientation (attitude must be considered as constant during individual measurement, within the required knowledge accuracy). There is no science driven need for an attitude control after the initial deployment-induced rotation are damped down. However ancillary requirements may impose attitude control, e.g., solar pointing for solar panel, or Earth pointing for ground transmissions. The swarm network concept doesn't require pointing for intra-swarm communication (omni-directional antenna).

Concerning, the global attitude of the swarm, there is no requirement for pointing or attitude control, but there is a scientific requirement on the knowledge of its absolute attitude in an absolute astrophysical frame.

\section{Technical Studies}

This section presents the results of the various studies conducted during the past year. 

\subsection{Measurements and Processing}
Each sensor-node (i.e., nanosatellite) is composed of three orthogonal electric dipoles (noted $h_1$, $h_2$, $h_3$) which provide a full polarimetric measurement of radio waves. Their physical length is set to $\sim$5m tip-to-tip \cite{Saks:2010uf,Boonstra:2016tc}. Each dipole is connected to a low noise amplifier (LNA) directly situated at the antenna feed point to reduce the impact of Radio Frequency Interference (RFI). This first amplification stage is critical in the whole amplification chain. The noise level of current designs is of the order of 3 to 5 nV/Hz$^{1/2}$. The LNA input impedance is large enough to limit the interaction between antenna and the satellite body to a capacitive coupling.

\begin{figure*}
\centering\includegraphics[width=0.8\linewidth]{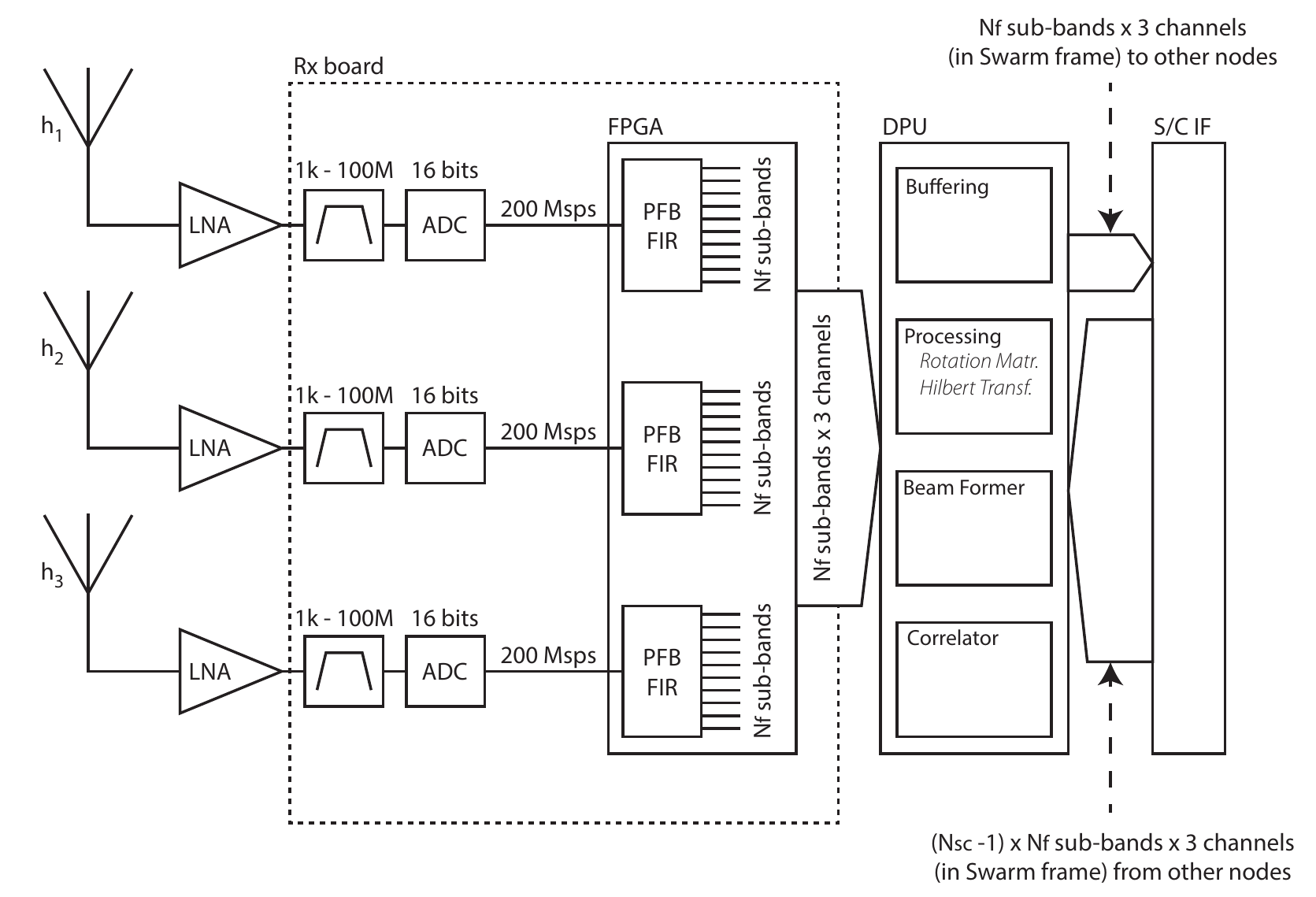}
\caption{Typical architecture for the acquisition and processing on a node. The measurement chain is composed of an antenna, a low noise amplifier (LNA) and a digitization front-end (filtering and conversion). The radio receiver (Rx board) is composed of those front-ends and an FPGA, which extracts $N_f$ sub-bands using polyphase filter banks (PFB) or impulse response filters (IRF). The data stream is sent to the digital processing unit (DPU), which manages all further processing. The data stream is also transmitted.}
\label{indivmeas}
\end{figure*}

Each measured and amplified electric signal is digitized by the radio receiver in the 1 kHz -- 100 MHz band. The three antennas are sampled simultaneously by parallel Analog-to-Digital converters (ADC). The rate of the digital signal is 200 Msamples per second. On-board Field Programmable Gate Arrays (FPGA) will produce a defined number of $N_f$ frequency channels (or sub-bands). A total of 3$N_f$ sub-bands will be produced by each sensor-node.
Fig.\ \ref{indivmeas} display the typical measurement chain on a single sensor-node.
This flux of raw data will then be processed jointly along with the similar flux coming from other sensor-nodes. Subsequent steps include data buffering, post-processing, beam forming and/or correlations.
The computing load to produce the array response represents a critical design for the instrument. 

% Distributed computing / Beamforming

\begin{figure*}
\centering\includegraphics[width=\linewidth]{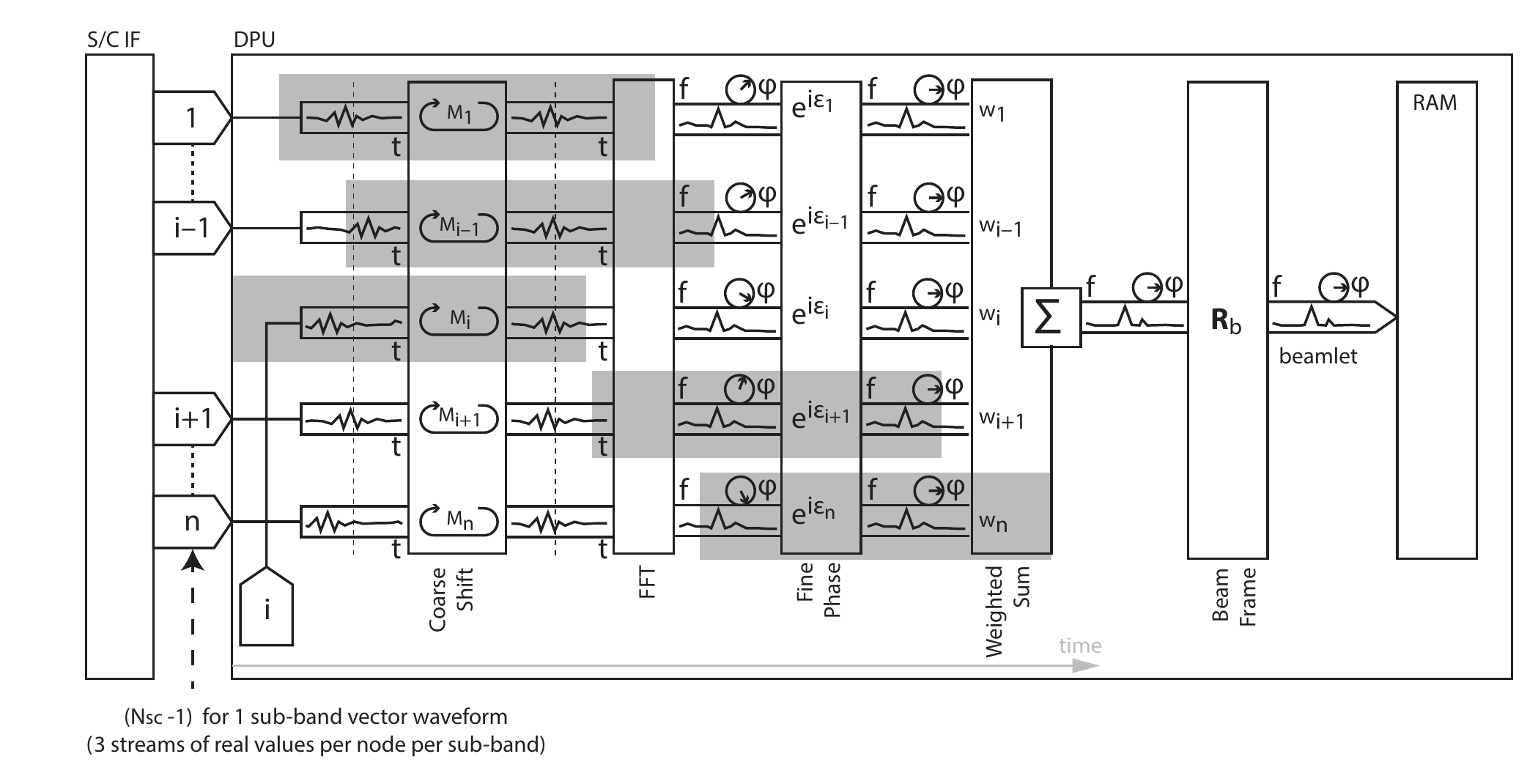}
\caption{Principle of Beam Forming done on a single node. The waveform from each node is processed the same way: coarse shifting using a shift register to correct for large scale radio path; an FFT to pass into the Fourier space; a fine shifting using a phase shift in Fourier space; and a weighted sum. As the data propagation path in the swarm network depends on the distance between the nodes, all waveforms are not processed simultaneously. The resulting three dimensional complex spectrum is rotated into the beam frame to obtain the complex beamlet spectrum.}
\label{distribmeas}
\end{figure*}

Assuming that all sensor signals have reached the central computation node, a dedicated post-processing unit is taking care of the beamforming/phase shifting of the signals.
Design studies have to tell if a central unit is necessary (e.g., a mother ship), or if the computation can be distributed among the nodes.
Fig. \ref{distribmeas} depicts a beamforming processing performed at the node $i$. After having received the $N_a-1$ time-tagged signals from the other nodes, the phasing step is performed in two steps: a coarse phasing step and a fine phasing step. The former is done by using a shift register to insert true-time delays (TTD) between the signals. Then, after undergoing a Fourier transform, complex phase coefficients are applied to each signal in the Fourier space. The coarse step helps inserting the necessary TTD to fold the remaining pointing phase to the $[0^\circ - 360^\circ]$ range.

The resulting three dimensional complex spectrum is rotated into the beam frame to obtain the complex beamlet spectrum. One can decide to allocate the responsibility of computing one beamlet to one node so that up to $N$ beamlets can be formed independently. One can add ``weights'' to the various nodes signal to shape the beam according to an arbitrary apodizing strategy. Classical weights are 1 to enable the beamforming of a theoretical array of identical antennas.
The phased signals can either be summed coherently (``beamformed'' mode) or be cross-correlated (``Interferometer'' mode) to compute the visibilities around the beamlet phase center.

\subsection{Orbitography}
Several orbital options have been studied in previous studied (Sun-Earth L1, Sun-Earth L2 or Lunar orbit). In the frame of NOIRE, a lunar equatorial orbit is investigated at this time. An Earth-Moon L2 orbit study was considered as an option at early stages of the study, but has not been further developed.  

The first step of the analysis is to evaluate the time of visibility for an earth observer for TM/TC (Telemetry and Telecommand) purposes as well as for scientific purposes (some science objectives requires to be hidden from Earth radio interferences). As the Moon is in the ecliptic plane, the location of the observer matters: with an observer at Earth's center (occultation of Earth), a study over a month is enough; with an observer at a specific location on Earth's surface, a study over a year is required. In all cases the sensitivity of the results with respect to the orbit altitude has to be evaluated. The lunar gravitational sphere of influence has a radius of 50000 km. A lunar orbit will then be inside that sphere. The altitude of the orbit is not constrained by the science requirements.

As a first step, we study a circular lunar orbit. We recall that the Moon radius is $\approx$1700 km. The period of orbits of semi major axes 5000 km or 10000 km are respectively $\approx$7.0 h and $\approx$24.9 h. The fraction of time in Earth occultation is inversely proportional to the distance. For the two orbital altitude previously given, the spacecraft is in Earth occultation condition 11\% and 5.5\% of the time, respectively. The case of elliptical orbits has been studied in order to try to increase the Earth occultation fraction time. However, the analysis allows us to rule out this option, as the orbit apsides are rotating in the Earth-Moon frame, leading to reduce the occultation time over course of a lunar revolution. 

Considering first a swarm with two nodes on the same orbit (with an offset), there are three possible effects depending on the offset (in-track, radial or off-plane). An in-track offset makes no specific effects, the nodes are just following one each other. A radial offset gives a relative rotation in the orbital plane. With off-plane offset, the nodes will have relative off-plane rotation. We can extend this simple scheme to a swarm with a larger number of nodes. In the simulation, 50 nodes are set randomly in an initial sphere of 100 km radius, and an orbital period for the swarm center is $\approx$9 h.
\begin{figure*}
\centering\includegraphics[width=0.49\linewidth]{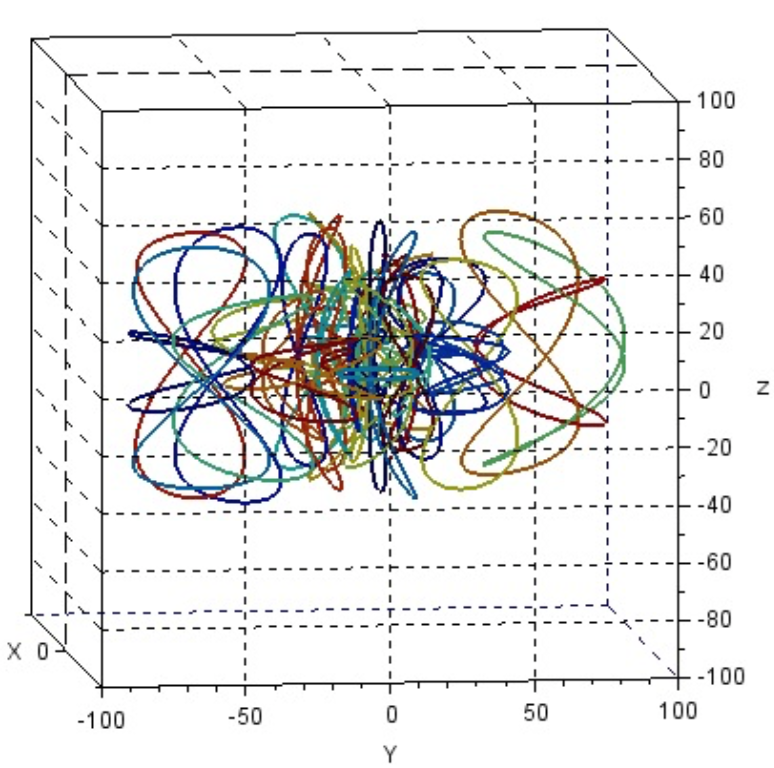}\includegraphics[width=0.49\linewidth]{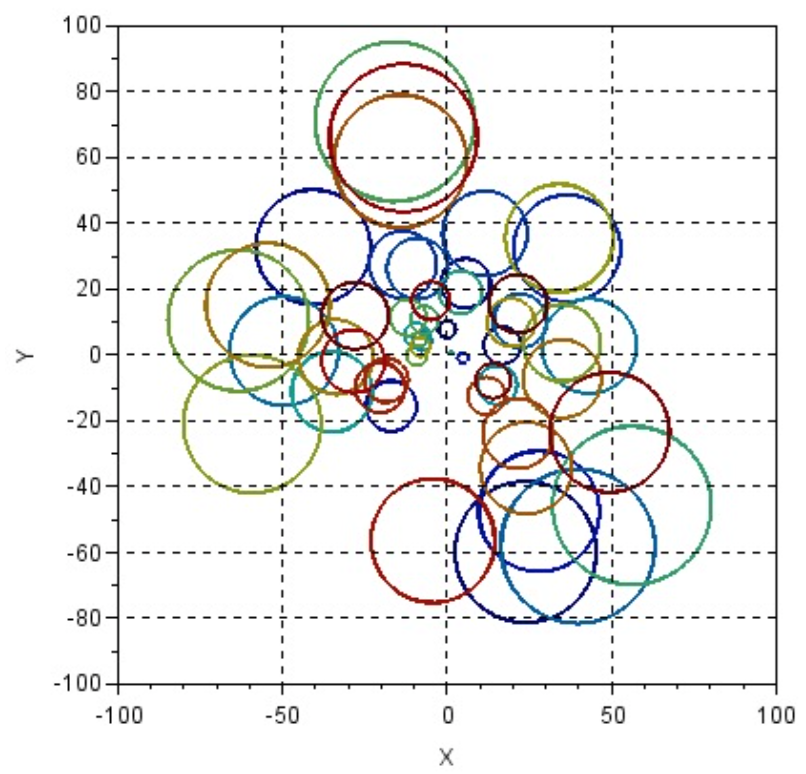}
\caption{Orbital simulation of a swarm of 50 elements, in circular Lunar orbit (orbital period of $\approx$9h). The left-hand panel shows the motions of each node in the center of mass frame of the swarm, over one orbital period. The right-hand panel shows the same data, but seen in the direction of the $z$-axis. Distances are given in km.}
\end{figure*}

The evolution of circular orbits with respect to gravitational perturbation (Earth, Sun, Lunar J2 gravitational moment) will be evaluated in further studies, as well as the swarm's deployment strategy.

\subsection{Location, Timing and Navigation}

\subsubsection{Location and time keeping}
Performances of the location and time system are directly constrained by the measurement requirements, as the swarm is the infrastructure support of measurement devices. As discussed in the previous sections, the first order estimate of the required precision on the node distances is of the order of 1 cm, and the time keeping precision is directly linked to the length of the individual snapshots to be combined. 

Based on the homogeneous swarm hypothesis, the initial selected location and timing concept is founded on GNSS-like strategy (Global Navigation Satellite System). Each node is both a transmitter and a receiver. By using a well-known TOA-CDMA technique (Time Of Arrival - Code Division Multiple Access), inter-satellites ranging could be precisely determined. Consequently, each node can use the rest of the swarm to localize itself.

The proposed concept is based on a S-band device with a channel for ranging determination and another one for science data. This concept was used for the proximity link between Rosetta and Philae, with 1 MHz for codes and 50 Hz for data. The model is based on the orbit modeling presented before and includes the relative velocities. In this first estimation, the assumptions are the following: the nodes don't have spin nor attitude control; they use omni-directional antenna both transmission and reception; there is no significant multipaths propagation; there is no electromagnetic interference in transmission-band; the swarm is composed of 50 nodes, all in operations; and the noise temperature doesn't depend on the antenna pattern (two extreme temperatures were considered: 30K in Lunar occultation, 300K concept). The simulation also assumes that clocks have no bias.

\begin{figure}
\includegraphics[width=\linewidth]{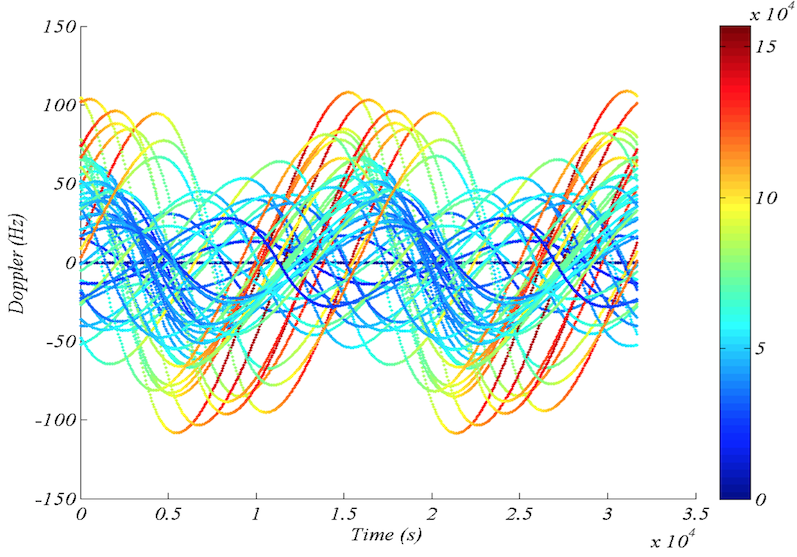}
\caption{Doppler shift (vertical axis) and node distance (color axis) plotted versus time for a simulation run.}
\label{fig:doppler-distance}
\end{figure}

\begin{figure}
\includegraphics[width=\linewidth]{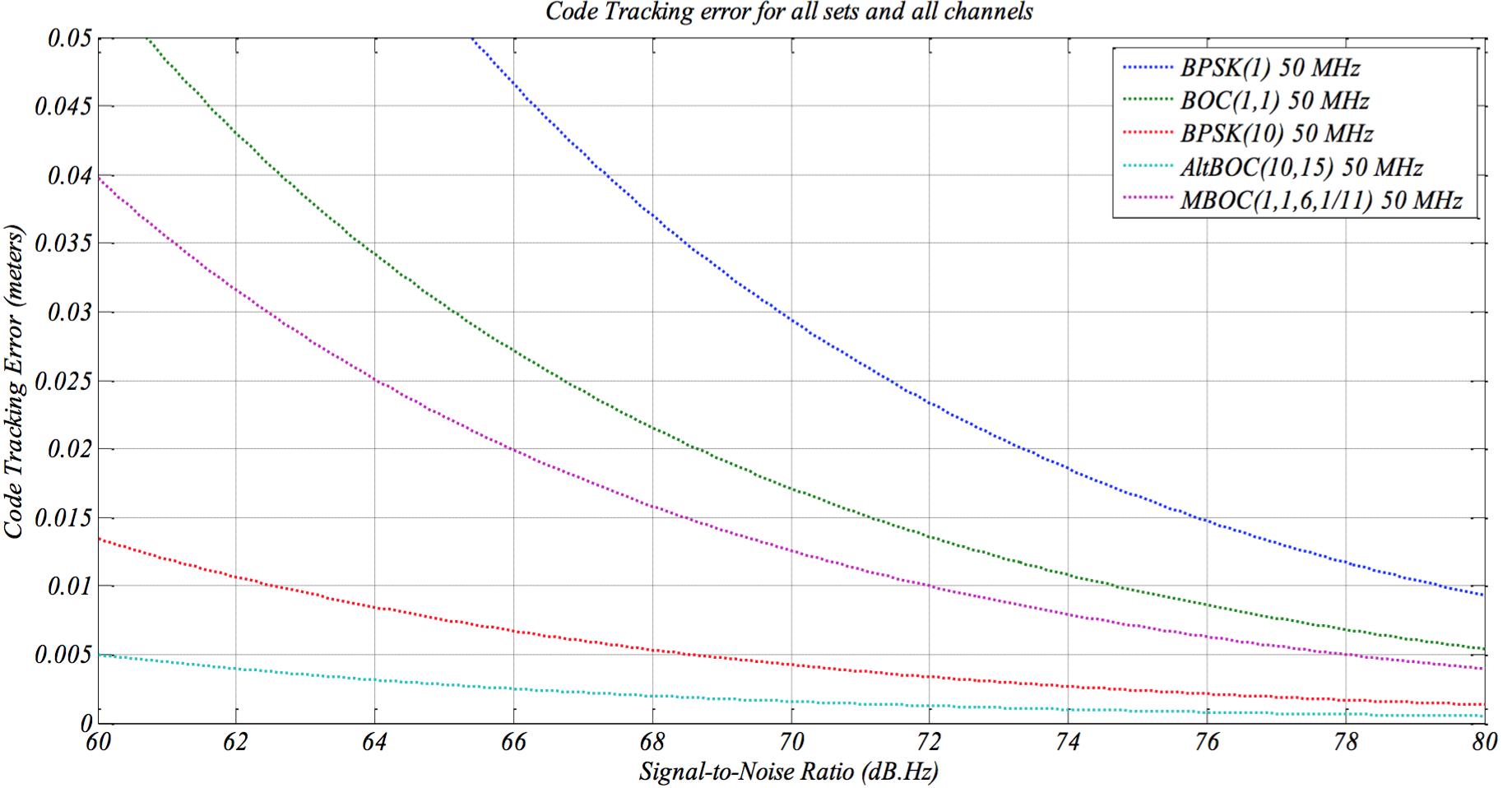}
\caption{Code tracking error as a function of Signal to noise ratio, for all sets}
\label{fig:code-tracking-error}
\end{figure}

The signal observed by a node $i$ is modeled as follows:
$$
\begin{array}{l}
P^{i1}_{F1,C1}=\rho^{i1}+h_i+h_1+c.(b^{i}_{C1}-b^{s1}_{C1})+\eta_1(F1,C1)\\
P^{i1}_{F2,C2}=\rho^{i1}+h_i+h_1+c.(b^{i}_{C2}-b^{s1}_{C2})+\eta_1(F2,C2)\\
\quad\quad\quad\,\,\,\vdots\\
P^{iN}_{F1,C1}=\rho^{iN}+h_i+h_N+c.(b^{i}_{C1}-b^{sN}_{C1})+\eta_N(F1,C1)\\
P^{iN}_{F2,C2}=\rho^{iN}+h_i+h_N+c.(b^{i}_{C2}-b^{sN}_{C2})+\eta_N(F1,C2)\end{array}
$$
where $P^{ij}_{F1,C1}$ and $P^{ij}_{F2,C2}$ are the pseudo distances between nodes $i$ and $j$ measured on frequency links $F1$ and $F2$ and for the codes $C1$ et $C2$; $\rho_{ij}$ the geometrical distance between nodes $i$ and $j$; $h_i$ is the clock error on node $i$; $\Delta\tau^{ij}=(b^{i}_{C1}-b^{sj}_{C1})$ is the instrumental delays for nodes $i$ and $j$ for the code C1; and $\eta_1(F1,C1)$ is the DLL (Delay Lock Loop) thermal noise tracking error on node $i$ for the frequency F1 and code C1, and $c$ the vacuum speed of light. As shown on Figure \ref{fig:doppler-distance}, the doppler shifts are rather small for this orbital configuration which makes frequency tracking easier. A detailed analysis of link budget also shows that the signal to noise ratio (C/N0, Carrier to noise density) is never going below 60 dB.Hz, which is a very favorable situation. Indeed, on simulated orbits, satellite ranging never exceed 180 km and with a EIRP (Effective Isotropic Radiated Power) of 1W, the link budget is excellent. In this way, the error estimation variance on the pseudo-distance tracking clearly decreased compared to Earth application. The Figure \ref{fig:code-tracking-error}, highlights the DLL standard deviation error in function of C/N0. The curves depend on several DLL tracking parameters (bandwidth filter, time integration, chip spacing, discriminator, \ldots) and, above all, depend on chip frequency rate and type of modulation.

For instance, by using faster modulation such as BPSK(10) (Binary Phase-Shift Keying modulation technique), the ranging accuracy of 1 cm in node to node could be easily achievable. The final accuracy of location not only depends on ranging accuracy but also on a factor called "Dilution of Precision" (DoP), which depends on the location of reference nodes compared to the receiving node. The simulation shows that $DoP<1$, which implies that by only considering thermal noise effects, the geometrical accuracy of 1 cm compared to a reference point of the swarm is achievable. Another challenging problem to cope with is not in theoretical aspects of localization strategy but specifically about the quality of calibration done on instrumental hardware biases. Moreover, the clock biases should be accurately estimated in order to maintain a global synchronization of all node local oscillators thanks to a time scale algorithm. This item will be the subject of a further study.

The distances between each pair of nodes are the interferometric baselines. This information must then be known on the nodes that combine the measurements into beamforming or interferometric products.

\subsubsection{Swarm shape and orientation}

In order to locate the source of the radio astronomical signals, or to achieve beamforming-based observations, the knowledge of the orientation of the baselines (and of the whole swarm) is also necessary. At first sight, it appears possible to reconstruct the 3D topology from the node-to-node distances. However multiple configurations  that generate the same history of node-to-node distance measurements can be found for a swarm of free drifting satellites. Removing the ambiguity implies therefore some additional observables that can be brought either  by the orbit determination of some of the nodes  or indirectly by the application of maneuvers. 
 
In this study, a ``minimalist'' navigation concept has been proposed to estimate the swarm's shape and orientation. 
It requires the knowledge of the position of at least one node, and reconstructs the absolute state of all the other nodes through the integration of the swarm dynamics and the processing of relative distance measurements. The efficiency of this concept is strongly dependent on the precision of the time and location system, but also the accuracy with which the determination of the reference node's position can be achieved with ``ordinary'' ranging and doppler measurements (such as those realized on TM/TC links for deep space probes) and associated on-board orbit extrapolation.

Various GNC (Guidance, Navigation and Control) strategies have been identified in order to optimise the accuracy of this navigation concept and a prototyping of the aforementioned navigation technique has been performed \cite{delpech_2017}. Complementary work is needed to address critical issues: the convergence of the estimator initialization, the design of an optimized and reconfigurable distributed estimator to reduce the computational load. These will be the subject of further studies.

\subsection{On-board Processing and Avionics}

In the growing market of nanosatellites, some low-cost low-power hardware solutions have been identified to be good candidates, based on COTS SoC (System on Chip). This kind of new component provides high processing power and flexibility thanks to multiple hardware architecture available (GPU, FPGA, multicore CPU). Even if some references are announced to be Rad-Tolerant, such solutions and associated architectures need to be evaluated from the EMC point of view (see below) and the mitigation of their potential impacts on the radio measurement carefully analyzed.

As regards the on-board processing, two candidate distribution schemes have been identified on the basis of well-known ground solutions: task distributed parallelism (such as Apache Mesos) versus data parallelism (such as MPI). Although these distribution schemes will have to be more deeply assessed, it already appears that a dedicated work has to be done on algorithm implementation, in order to relax the functional and QoS requirements applicable to the inter-node data communication system, which is expected to be challenging and costly. Further studies and trade-offs will be considered, as soon as some key processing parameters will be more precisely known: elementary measurement duration, number of successive or integrated acquisitions, data volume to be acquired and exchanged, sequencing of local and distributed processing.

\subsection{Electro-Magnetic Compatibility}
Radio receivers mounted on the NOIRE swarm nodes have to measure low level electric fields. These receivers are very sensitive. Electromagnetic compatibility (EMC) requirements are thus one of the main constraints on the system. In order to ensure electromagnetic cleanliness, EMC concepts must be taken into account in early design phases. An EMC work group shall be set up as soon as possible during implementation. This working group shall be composed of: EMC experts, instrument team members and system representatives. The EMC working group will set up a precise EMC control plan and update it throughout the project (including during design and test phases). The EMC control plan contains studies and analyses proving that instruments and system are compliant with EMC requirements, design rules, frequency control plan, as well as test concepts and procedures. The EMC working group will also assess any deviation and exception to the EMC control plan and and propose adequate recommendations or waivers. 

The requirements defined in ECSS-E-ST-20C (Electrical and electronic) and ECSS-ST-20-07C (ElectroMagnetic compatibility) are applicable. These requirements will be reinforced according to the mission-specific constraints. EMC control plans from the Solar Orbiter or JUICE mission can be used as a good starting point. However they will have to be updated taking into account the specific NOIRE concept platform (swarm of small interconnected nanosatellites).

\subsection{Communications and Networking}

\subsubsection{Inter-node communication}
The proposed inter-satellite link (ISL) is based on omni-directional communication to avoid implementing pointing systems between nodes. The studied networking scheme is a physical connectivity limited to the nearest neighbors (in order to limit the power consumption of the telecom system), the communication at swarm scale being achieved through successive hops by means of network protocols. As said before, such an ISL would be compatible  with the auto-location system and share the same physical link. Further dimensioning of the hardware equipment and associated link budget computation need data rates to be estimated for each function (location and navigation, time keeping protocol, distributed processing, command and control, etc.).

\subsubsection{Networking paradigm}
The orbital 3D modeling of the swarm a strongly dynamical topology. The relative locations of each node is very variable in time. This will challenge the network routing protocol, as the route joining two nodes is not persistent. Node are able to communicate with the surrounding nodes, within the effective range of their communication antenna. Several hops from nodes to nodes may be needed to route a packet to its destination. The node mobility and the power limitation must be taken into account to select or design suitable routing protocols. Such type of information routing problematic has been studied for \textsl{Mobile Ad hoc NETworks} (MANET) without infrastructure where all stations are mobile. The IETF (Internet Engineering Task Force) is standardizing routing protocols for MANET \cite{rfc2501}.

A MANET is characterized by the following properties. Nodes are mobile units, which are moving freely and arbitrarily. The network topology can thus be modified anytime, in a unpredictable manner. The topological links may be mono- or bidirectional. The units are autonomous and may have limited power. The power resources management has be to taken into account. There isn't any preexisting nor centralized infrastructure. All nodes communicate through there own wireless interfaces. As there is no infrastructure, all nodes are routers and participate to the discovery of routes between nodes. The mobiles nodes are thus forming an ad-hoc network infrastructure. The way the network topology is changing is a dimensioning characteristic of the network, as it implies dynamical reconfiguration of the network and may result into frequent node disconnections. Several classifications of MANET routing algorithms have been proposed (see, e.g., \cite{Saeed:2012es} and \cite{sharma:2016ub}). The latter reference defines three classes of MANET: proactive, reactive and hybrid networks. Proactive MANET are implementing a knowledge of the network topology in each node, which uses a large bandwidth overhead to maintain this knowledge. On reactive MANET connexions are established on demand from peer to peer and there is no need to maintain a routing table. However, establishing a route takes a longer time (not predictable) than with a proactive network. Hybrid MANET are trying take advantage of both solutions. For instance the ZRP (Zone Routing Protocol) a proactive scheme is used between close neighbors and a reactive scheme is used for further connections. In the case of NOIRE, setting up a MANET seems appropriate, probably using ZRP scheme, as we want to limit the system bandwidth compared to the science bandwidth.

\subsubsection{Communication with Earth}
Direct communication to Earth from a lunar nanosatellite is of course challenging, as long as a significant data rate is needed. Except the relay solution (the carrier that has deployed the swarm is still in the Moon's vicinity and could be used for this purpose), the swarm itself could be envisaged as a synthetic antenna for direct-to-Earth communication. This original concept has not yet been studied further.

\subsection{Satellite Architecture}

No traditional architecture studies (sizing, layout, subsystems dimensioning) have be conducted, but the first identified needs for capacity or performance, combined to the increasing offer of nanosatellite equipment, e.g., energy and power conditioning, attitude control sensors and actuators, structure elements, launcher interface, give confidence in the availability of hardware solutions to implement the swarm's satellite.

Nano-satellites or very small platforms tend to be strategic for spatial applications. Guiding and controlling very small satellite trajectories as well as their orbital drift ask for compact, efficient, and robust propulsion systems. However, traditional propulsion systems hardly match the required constraints for this new generation of small satellites. Most of these technologies have been optimized for operation in a range of power that serves the needs of usual space missions (North South Station Keeping and Orbit raising). For example, a relatively large number of Hall Effect thrusters ranges from almost 200 W in power level to several tens of kilowatts.  New innovative and efficient propulsion systems are then needed, and new tracks have to be explored

%If further orbit studies reveal a need for station acquisition and keeping,  micro-propulsion solutions will need to be carefully studied, in terms of capacity and performance, but also in terms of mass, volume and power budgets.

\section{Conclusions and Perspectives}

As this stage, the NOIRE study has began preliminary work to identify the main constraining parameters of the proposed concept, along with system-level enabling solutions. If the feasibility of such a system has not yet been demonstrated, no strong show-stopper is identified. Thus, the concept is not yet ready to be implemented, and further studies are identified in the following areas.

The study has highlighted the need for a measurement system modeling, jointly with scientific measurement modeling and system modeling, in order to  refine the system dimensioning and establish realistic requirements. The NOIRE concept is based on the principle of a system-wide instrument, so that the measurement equation requires a mixed top-down (from measurement performance to requirements on the system) and bottom-up approach (incorporating device performance into noise budgets). This will contribute to a better definition of the required on-board processing, and thus give cross-inputs with hardware and software dimensioning and benchmarking activities.

The concepts of autonomous location and time keeping inside nanosatellite swarm need to be further assessed through the development of an autonomous method for an on-board implementation. A network-based (no master clock, time lapse dissemination) time keeping strategy is foreseen, allied with "GNSS-less" location techniques.

Orbitography studies are necessary to assess the long term effects of gravitational perturbations on the swarm topology, and the associated impacts on scientific measurement capabilities. The initial conditions should also be analyzed, in relation with the possible ways of deploying and "shaping" the swarm.

Last but not least, GNC studies would benefit from location and orbitography studies to push forward a novel absolute navigation concept.

\appendices
\section{The NOIRE team}
\label{app:team}
The NOIRE team is composed of research scientists and engineers from several laboratories and institutes in France. The core laboratories involved are the following (naming the people who participated in the study): 
\begin{itemize}
\item\textsl{LESIA, Obs. Paris, CNRS, PSL, France} --- B.~Cecconi, C.~Briand, P.~Zarka, L.~Lamy, M.~Moncuquet,
 M.~Maksimovic, R.~Mohellebi, A.~Zaslavsky, Y.~Hello, B.~Mosser, B.~Segret, S.~Chaintreuil. 
\item\textsl{APC, Univ.\ Paris 7 Denis Diderot, France} --- M.~Agnan, M.~Bucher, Y.~Giraud-Heraud, 
H.~Halloin, S.~Katsanevas, S.~Loucatos, G.~Patanchon, A.~Petiteau, A.~Tartari.
\item\textsl{LUPM, Univ.\ Montpellier, France} --- D.~Puy, E.~Nuss, G.~Vasileiadis.
\end{itemize}
In addition to the research laboratories, a team of engineers from CNES has been actively involved: Andr\'e Laurens, Alain Lamy, David Valat
Franck Barbiero, Jean-Jacques Metge, Michel Delpech, Mickael Bruno, Patrick G\'elard. Others persons from CNES have been associated to discussions: 
Antoine Basset, Claude Boniface, C\'eline C\'enac-Morth\'e, C\'eline Loisel, Cl\'ement Dudal, Johan Panh, Marie-France Del Castillo, Roberto Camarero and Pierre-Marie Brunet.

Other laboratories have been involved, with significant contribution:
\begin{itemize}
\item\textsl{CEA/SAp/IRFU, Univ.\ Paris 7, Saclay, France} --- J.~Girard.
\item\textsl{ONERA/Toulouse, France} --- A.~Sicard,  Q.~Nenon.
\item\textsl{IRAP, Toulouse, France} --- M.~Giard.
\item\textsl{GEPI, CNRS-Obs. de Paris, France} --- C.~Tasse.
\item{LPC2E, CNRS-Univ.\ d'Orl\'eans, France} --- J.-L.~Pin\c con, T.~Dudok de Wit, J.-M.~Grie{\ss}meier. 
\item\textsl{C2S/TelecomParis, France} --- P.~Loumeau, H.~Petit, T.~Graba, R.~Mohellebi, P.~Desgreys, Y.~Gargouri, C.~Jabbour.
\item\textsl{IMAG, Grenoble, France} --- S.~Mancini.
\end{itemize}
In addition, so-called space campuses (university nanosatellite groups) were involved:
\begin{itemize}
\item\textsl{Centre Spatial Universitaire de Montpellier-N\^imes, Universit\'e de Montpellier} --- L.~Dusseau. 
\item\textsl{Fondation Van Allen, Institut d'\'Electronique du Sud, Universit\'e de Montpellier} --- F.~Saign\'e.
\item\textsl{Campus Spatial Diderot, UnivEarthS, Sorbonne Paris Cit\'e} --- M.~Agnan. 
\item\textsl{C$\!\;^2\!$ERES, ESEP/PSL} --- B.~Mosser, B.~Segret. 
\end{itemize}
International collaboration was also initiated with the Dutch low frequency radio astronomy team (ASTRON, TU-Eindhoven, Radboud Univ.\ Nijmegen, TU-Delft, TU-Twente, ISISpace, Hyperion Technologies): W.~Baan, M.~Bentum, A.-J.~Boonstra, S.~Engelen, M.~Klein-Wolt, B.~Monna, H.~Pourshaghaghi, J.~Rotteveel, P.~K.~A van Vugt.

\bibliographystyle{IEEEtran}
\bibliography{IEEEabrv,noire}

\thebiography
\begin{biographywithpic}
{Baptiste Cecconi}{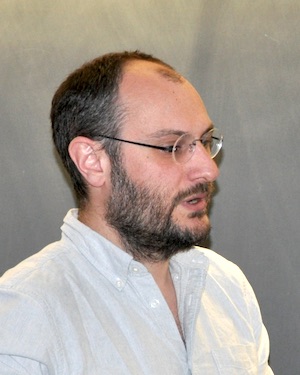}
received his Ph.D.\ in astrophysics from Univ.\ Paris Diderot (Paris 7) in 2004. He is currently a research scientist at LESIA, Observatoire de Paris. His current research activities include the study of planetary magnetospheres using radioastronomy. He is involved in many instrumental (present and future, ground-based and space-borne) radio astronomy projects. He is the Science lead of the NOIRE study. 
\end{biographywithpic}

\begin{biographywithpic}
{Julien N. Girard}{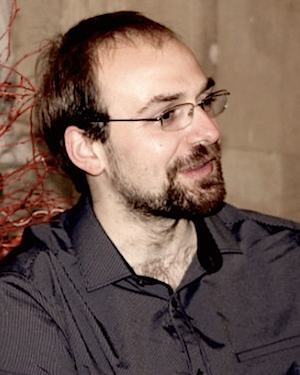}
received his Ph.D.\ in astrophysics from Observatoire de Paris in 2013. He joined CEA/AIM as an associate professor of Univ.\ Paris Diderot (Paris 7). His current research activities include the study of radio transients with ground-based SKA-class radio telescopes (LOFAR and NenuFAR). He is also involved in instrumental development and commissioning for low frequency radioastronomy. 
\end{biographywithpic}

\begin{biographywithpic}
{Andr\'e Laurens}{bio-laurens}
joined CNES in the early 1980s and has contributed to many space missions, mainly working in satellite image processing, spacecraft control and command, balloonborne astrophysics and atmospheric chemistry experiments. He is currently a space systems expert in PASO, the CNES Early Mission Studies team, and coordinator of Space Sciences studies. He is involved in many future mission studies conducted by CNES in this area.\end{biographywithpic}

\begin{biographywithpic}
{Carine Briand}{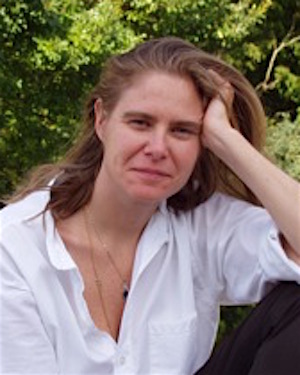} is presently research scientist at LESIA, Observatoire de Paris.  Her current research activities include particle acceleration and energy conversion in space and laboratory plasmas. She is involved in several Space Weather activities and groups. She also participates in the commissioning activities of NenuFAR, a low frequency ground based radio interferometer.   
\end{biographywithpic}

\begin{biographywithpic}
{Martin Bucher}{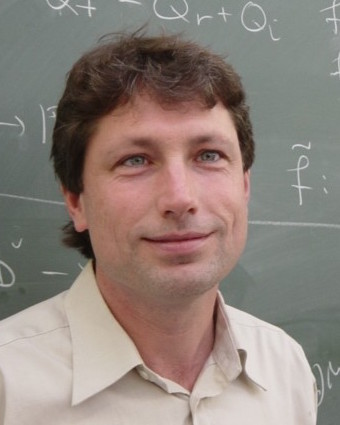}
 is presently Directeur de recherche at the CNRS and works at the ``Laboratoire Astroparticules et Cosmologie'' (APC), Univ.\ Paris Diderot (Paris 7). Born in Germany, he attended Yale, University of California (A.B. Physics 1986) and Caltech (Ph.D 1990). Before coming to France, Bucher has held positions at the Institute for Advanced Study, Princeton University, the Yang Institute for Theoretical Physics at Stony Brook, and the University of Cambridge. His current research activities include theoretical cosmology, particle theory, CMB studies. 
\end{biographywithpic}

\begin{biographywithpic}{Moustapha Dekkali}{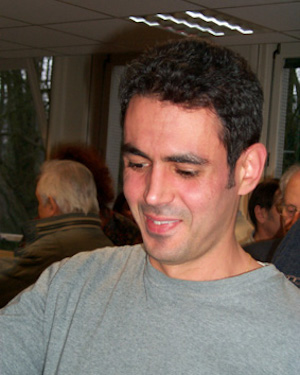}
is a radio frequency engineer at LESIA, Observatoire de Paris. He has been actively developing and building low frequency radio instruments for space missions, such as Solar Orbiter/RPW and Bepi-Colombo/MMO/Sorbet. He is working on R\&D studies for future generation of space radio instrumentation. 
\end{biographywithpic}

\begin{biographywithpic}{Quentin Nenon}{bio-nenon}
is Ph.D.\ student at ONERA, in Toulouse, France. He is working on the modeling of planetary radiation belts.\\[4em]
\end{biographywithpic}

\begin{biographywithpic}{Jean-Mathias Grie{\ss}meier}{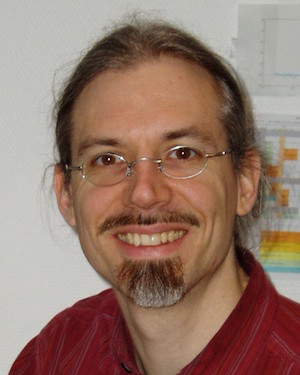}
received his Ph.D. from the Technical University of Braunschweig in 2006. He is currently a research scientist at LPC2E Orl\'eans and the Nan\c cay Radio Station. His current research activities include the search for radio emission from extrasolar planets and the study of pulsars at low radio frequencies ($<$200 MHz). He is also strongly involved in LOFAR and NenuFAR, two ground-based radio telescopes at low frequencies.
\end{biographywithpic}

\begin{biographywithpic}
{Boris Segret}{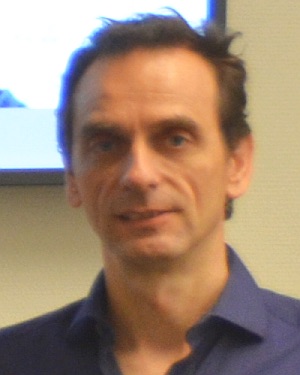} received his first M.S. degree in Aerospace Engineering from ISAE-ENSMA French National College in 1991 and his second M.S. deree in Space Instrumentation from Paris Observatory in 2011. He works as a research engineer for ESEP (Exploration Spatiale des Environnements Planetaires) at Paris Observatory, using nanosatellites as a means for technology development and new scientific applications.
\end{biographywithpic}

%\begin{biographywithpic}
%{C\'eline Loisel}{bio-loisel}
%graduated the French Engineer School INT Telecom in 2004 with a specialization in Space Telecommunications. 
%She has been working at the French Space Agency (CNES) since 2004 as a Space Telecommunication engineer. Specialized in telemetry and telecommand transmission since 2007, she is responsible of the S-band ground to board link definition and performances of scientific missions (Jason series, SWOT, SVOM). Celine works also for the Earth observation satellites developed by CNES by evaluating their link performances and compatibility with the ground stations. She is finally involved in Deep Space studies, operations and future inter-satellite links like Rosetta.\end{biographywithpic}

%\begin{biographywithpic}
%{Antoine Basset}{bio-basset.jpeg}
%is a software engineer and holds a Ph.D.\ in signal processing since 2015, which was applied to biomedical imaging.
%Now with CNES, he helps French laboratories developing and integrating the Euclid science ground segment, and also develops system tools for it.
%He mainly focusses on image processing and high performance computing.\end{biographywithpic}

\begin{biography}{Alain Lamy}
is senior expert in spaceflight dynamics, and for this reason, is involved in many space mission studies, either in Space Science and Earth Observation.\\[3em]
\end{biography}

%\begin{biography}{Claude Boniface}
%is a space propulsion engineer, specialized in electric propulsion subsystems.
%\end{biography}

%\begin{biography}{C\'eline C\'enac-Morth\'e} is electrical engineer, involved in satellite energy subsystem design, with a special interest in battery technologies.\end{biography}

\begin{biographywithpic}{David Valat}{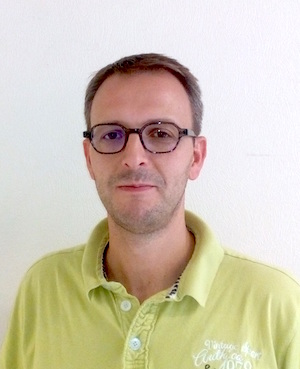} 
received his M.Sc. degree in 1997 in metrology and signal processing. He joined the Paris Observatory, SYRTE department as an R\&D engineer. For 12 years, he has been involved in time and frequency transfer and time scales activities. In 2011, he joined the French space agency CNES as time and frequency expert. His main interests are characterization of space grade oscillators, clocks remote comparison by GNSS/optical links, absolute calibration of GNSS reception chains and time scales generation.
\end{biographywithpic}

%\begin{biography}{Franck Barbiero} is a radiofrequency engineer, involved in GNSS signal processing and design of location and navigation systems.\end{biography}

%\begin{biographywithpic}{Jean-Jacques Metge}{bio-JJMetge.jpg} is expert in on-board software, mainly involved in software development for Space Science experiments. He was previously one of the designers of the new distributed avionics platform (Integrated Modular Avionics) embedded on board of the new generation of the Airbus planes family. \end{biographywithpic}

%\begin{biography}{Johan Panh} is CEM senior expert in CNES, and for this reason has contributed to many space missions design, particularly in the area of science experiments.\end{biography}

%\begin{biography}{Marie-France Del Castillo} is a cost estimation expert in CNES Procurement, Sales and Legal Affairs Directorate\end{biography}

\begin{biographywithpic}{Michel Delpech}{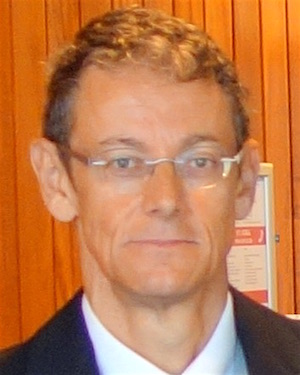} received his Ph.D. in automation \& Robotics from Univ. Toulouse/ISAE in 1985 and joined CNES in 1988 as robotics engineer. Strongly involved in formation flying since 2000, he was technical lead of the french contribution to the PRISMA mission. He is presently Senior Expert in Guidance Navigation and Control at CNES. His current activities include R\&D work on various GNC topics (spacecraft rendezvous, landing on small bodies, planetary rover navigation), participation to mission concept studies focused on formation flying, supervision of H2020 strategic research cluster in Space Robotics. \end{biographywithpic}

\begin{biographywithpic}
{Mickael Bruno}{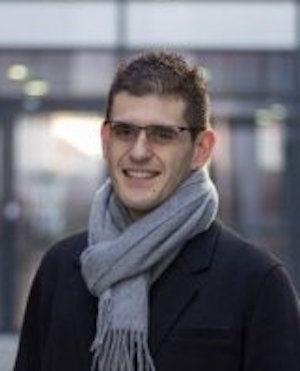} 
graduated the French Engineer School ENSEEIHT in 2015 with a specialization in electronics and signal processing. He joined the French Space Agency (CNES) in 2016. He is an on-board electronics and signal processing engineer, mainly involved in satellite payload design.\\[1em]
\end{biographywithpic}

%\begin{biography}{Roberto Camarero} is an on-board electronics and signal processing engineer, mainly involved in satellite payload design.\end{biography}

\begin{biographywithpic}{Patrick G\'elard}{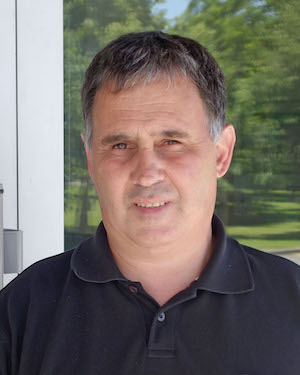} graduated with a Master of Advanced Studies in Artificial intelligence and  networking. He has 30 year's experience in networking and telecommunication technologies. He joined the CNES in 1990 as protocol engineering network expert. He has been involved in multiple projects in satellite communications. Now for eight years he has been in charge of research axis ``terrestrial and satellite infrastructure convergence''. This has been led him to work on subject R\&D studies like ``Software Defined Networking and Virtualization for Broadband Satellite Networks'', ``Machine learning for network traffic classification and prediction'', ``Satellite IoT and M2M Network''.\end{biographywithpic}

\begin{biographywithpic}{Alber-Jan Boonstra}{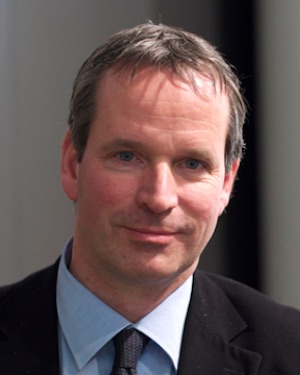} 
was born in The Netherlands in 1961. He received the B.Sc. and M.Sc. degrees in applied physics from Groningen University, Groningen, the Netherlands, in 1984 and 1987, respectively. In 2005 he received the PhD degree for his thesis Radio frequency interference mitigation in radio astronomy from the Delft University of Technology, Delft, The Netherlands. He was with the Laboratory for Space Research, Groningen, from 1987 to 1991, where he was involved in developing the short wavelength spectrometer (SWS) for the infrared space observatory satellite (ISO). In 1992, he joined ASTRON, the Netherlands Foundation for Research in Astronomy, initially at the Radio Observatory Westerbork, Westerbork, The Netherlands. He is currently with the ASTRON R\&D Department, Dwingeloo, The Netherlands, where he heads the DSP group. His research interests lie in the area of signal processing, specifically RFI mitigation by digital filtering.\end{biographywithpic}

\begin{biographywithpic}{Mark J. Bentum}{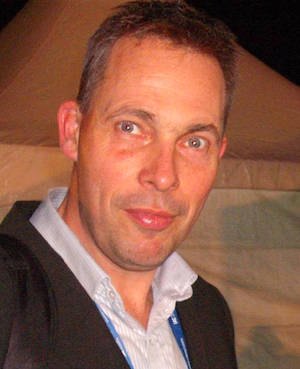}
(S'92, M'95, SM'09) was born in Smilde, The Netherlands, in 1967. He received the MSc degree in Electrical Engineering 
(with honors) from the University of Twente, Enschede, The Netherlands, in August 1991. In December 1995 he received 
the PhD degree for his thesis ÓInteractive Visualization of Volume DataÓ also from the University of Twente. From 
December 1995 to June 1996 he was a research assistant at the University of Twente in the field of signal processing 
for mobile telecommunications and medical data processing. In June 1996 he joined the Netherlands Foundation for 
Research in Astronomy (ASTRON). He was in various positions at ASTRON. In 2005 he was involved in the eSMA project 
in Hawaii to correlate the Dutch JCMT mm-telescope with the Submillimeter Array (SMA) of Harvard University. From 
2005 to 2008 he was responsible for the construction of the first software radio telescope in the world, LOFAR 
(Low Frequency Array). In 2008 he became an Associate Professor in the Telecommunication Engineering Group at the 
University of Twente. From December 2013 till September 2017 he was also the program director of Electrical 
Engineering at the University of Twente. In 2017 he became a Full Professor in Radio Science at Eindhoven 
University of Technology. He is now involved with research and education in radio science. His current research 
interests are radio astronomy, short-range radio communications, novel receiver technologies (for instance in the 
field of radio astronomy), channel modeling, interference mitigation, sensor networks and aerospace. Prof. Bentum 
is a Senior Member of the IEEE, Chairman of the Dutch URSI committee, vice chair of the IEEE Benelux section, 
initiator and chair of the IEEE Benelux AES/GRSS chapter, and has acted as a reviewer for various conferences and 
 journals. \end{biographywithpic}

%\begin{biography}{Pierre-Marie Brunet} is a computer architecture expert, involved in the development of space mission ground segments.\end{biography}

\end{document}